\documentclass{WileyMSP-template}
\usepackage{color} % for highlighting
\usepackage{soul} % for highlighting
\usepackage{url}
\usepackage{mhchem}

\begin{document}

\pagestyle{fancy}

\title{Chemical profiles of the oxides on tantalum in state of the art superconducting circuits}

\maketitle

\author{Russell A. McLellan$^\dag{}$}
\author{Aveek Dutta$^\dag{}$}
\author{Chenyu Zhou}
\author{Yichen Jia}
\author{Conan Weiland}
\author{Xin Gui}
\author{Alexander P. M. Place}
\author{Kevin D. Crowley}
\author{Xuan Hoang Le}
\author{Trisha Madhavan}
\author{Youqi Gang}
\author{Lukas Baker}
\author{Ashley R. Head}
\author{Iradwikanari Waluyo}
\author{Ruoshui Li}
\author{Kim Kisslinger}
\author{Adrian Hunt}
\author{Ignace Jarrige}
\author{Stephen A. Lyon}
\author{Andi M. Barbour}
\author{Robert J. Cava}
\author{Andrew A. Houck}
\author{Steven L. Hulbert}
\author{Mingzhao Liu*}
\author{Andrew L. Walter*}
\author{Nathalie P. de Leon*}

$^\dag{}$These authors contributed equally.

% Dedication

\dedication{}

\begin{affiliations}
R. A. McLellan, Dr. A. Dutta, Dr. A. P. M. Place, X. H. Le, T. Madhavan, Y. Gang, Prof. S. A. Lyon, Prof. A. A. Houck, Prof. N. P. de Leon\\
Department of Electrical and Computer Engineering, Princeton University \\
Princeton NJ USA\\
Email Address: npdeleon@princeton.edu
\medskip

Dr. C. Weiland\\
ORCID: 0000-0001-6808-1941 \\
Materials Measurement Science Division \\ Material Measurement Laboratory\\
National Institute of Standards and Technology \\
Gaithersburg MD USA

\medskip

Dr. X. Gui, Prof. R. J. Cava \\
Department of Chemistry, Princeton University \\
Princeton NJ USA\\
\medskip

K. D. Crowley \\
Department of Physics, Princeton University \\
Princeton NJ USA\\
\medskip

Dr. I. Waluyo, Dr. A. Hunt, Dr. Ignace Jarrige, Dr. A. M. Barbour, Dr. S. L. Hulbert, Dr. A. L. Walter\\
National Synchrotron Light Source II\\
Brookhaven National Laboratory\\
Upton, NY, USA\\
Email Address: awalter@bnl.gov
\medskip

Dr. C. Zhou, Dr. Y. Jia, L. Baker, Dr. A. R. Head, R. Li, K. Kisslinger, Dr. M. Liu\\
Center for Functional Nanomaterials\\
Brookhaven National Laboratory\\
Bldg. 735\\
P.O. Box 5000\\
Upton, NY 11973-5000\\
Email Address: mzliu@bnl.gov
\end{affiliations}

\keywords{tantalum, X-ray photoelectron spectroscopy, oxide, dielectric loss, superconducting thin films, qubits}

\newpage
\begin{abstract}

% Please insert your abstract here
Over the past decades, superconducting qubits have emerged as one of the leading hardware platforms for realizing a quantum processor. Consequently, researchers have made significant effort to understand the loss channels that limit the coherence times of superconducting qubits. A major source of loss has been attributed to two level systems that are present at the material interfaces. We recently showed that replacing the metal in the capacitor of a transmon with tantalum yields record relaxation and coherence times for superconducting qubits, motivating a detailed study of the tantalum surface. In this work, we study the chemical profile of the surface of tantalum films grown on c-plane sapphire using variable energy X-ray photoelectron spectroscopy (VEXPS). We identify the different oxidation states of tantalum that are present in the native oxide resulting from exposure to air, and we measure their distribution through the depth of the film. Furthermore, we show how the volume and depth distribution of these tantalum oxidation states can be altered by various chemical treatments. By correlating these measurements with detailed measurements of quantum devices, we can improve our understanding of the microscopic device losses.

\end{abstract}

\setlength{\parindent}{20pt}

\section{Introduction}

Superconducting qubits are the basis of many efforts to build large scale quantum computers, and have enabled key demonstrations of quantum algorithms \cite{kandala_hardware-efficient_2017}, quantum error correction \cite{corcoles_demonstration_2015, gong_experimental_2022, kelly_state_2015, reed_realization_2012, sivak_real-time_2022}, quantum many body physics \cite{mi_time-crystalline_2022, satzinger_realizing_2021, andersen_observation_2022, mi_noise-resilient_2022}, and quantum advantage in performing specific tasks \cite{arute_quantum_2019}. Despite this activity, little progress has been made in identifying and addressing the microscopic source of loss and noise in the constituent materials. The lifetimes of current 2D transmons are believed to be limited by microwave dielectric losses \cite{place_new_2021, wang_towards_2022}. Recent work has measured the dielectric loss tangent of high-purity bulk sapphire as $15(5)\times 10^{-9}$ at qubit operating conditions \cite{read_precision_2022}. This loss tangent would result in a qubit lifetime of several milliseconds if it were the only source of dielectric loss, suggesting that losses are dominated by uncontrolled defects at surfaces and interfaces or by material contaminants \cite{wang_surface_2015}. 

We have recently demonstrated that tantalum (Ta) based planar transmon qubits can exhibit record lifetimes over 0.3 ms and quality factors (Q) over 7 million \cite{place_new_2021}, exceeding the prior established record of niobium (Nb) and aluminum (Al) based planar transmon qubits by a factor of three. Other groups have recently realized Ta qubits with Q over 10 million and resonators with low power Q over 4 million \cite{wang_towards_2022, lozano_manufacturing_2022}. We hypothesize that one key advantage of Ta is its robust, stochiometric oxide, which is resistant to a wide range of aggressive chemical processes \cite{face_high_1986, face_nucleation_1987}. These observations motivate a careful study of the oxide species at the surface of Ta, as the amorphous oxide layer at the surface is a plausible source of dielectric loss arising from two-level systems \cite{muller_towards_2019, de_leon_materials_2021}. We have recently reported measurements on superconducting resonators showing that two-level systems at material interfaces are the dominant source of dielectric loss for tantalum devices, and that some of these two-level system defects reside in the surface oxides of tantalum \cite{kevin_resonator}. In that work, we estimated that the oxide layer is responsible for around half of the surface-related losses in state-of-the-art devices by using detailed comparisons between devices treated with different acids. In this work, we investigate the nature of the tantalum oxide resulting from these chemical treatments.

In this study, we use variable energy X-ray photoelectron spectroscopy (VEXPS) to characterize the surface of Ta thin films employed in state-of-the-art superconducting circuits, with the aim of identifying possible microscopic sources of loss and noise. We measure and analyze Ta4f ionizations with different peaks across a range of incident X-ray energies to vary the sampling depth and use the combined dataset to build a depth profile of the different oxide stoichiometries. We show that the Ta surface is dominated by the fully oxidized \ce{Ta2O5}, but that between the \ce{Ta2O5} surface layer and the bulk metal, suboxide species containing Ta$^{3+}$ and Ta$^{1+}$ are present. Compared to similar measurements of the oxides of Nb \cite{Anjali_Nb:2021}, the Ta oxide layer is thinner and the interfaces are more abrupt. We apply our measurement and analysis technique to films treated with different acid processes and show that it is possible to controllably grow and shrink the Ta oxide layers.

\section{Experiment}

\begin{figure}[h]
\centering
  \includegraphics[width=5in]{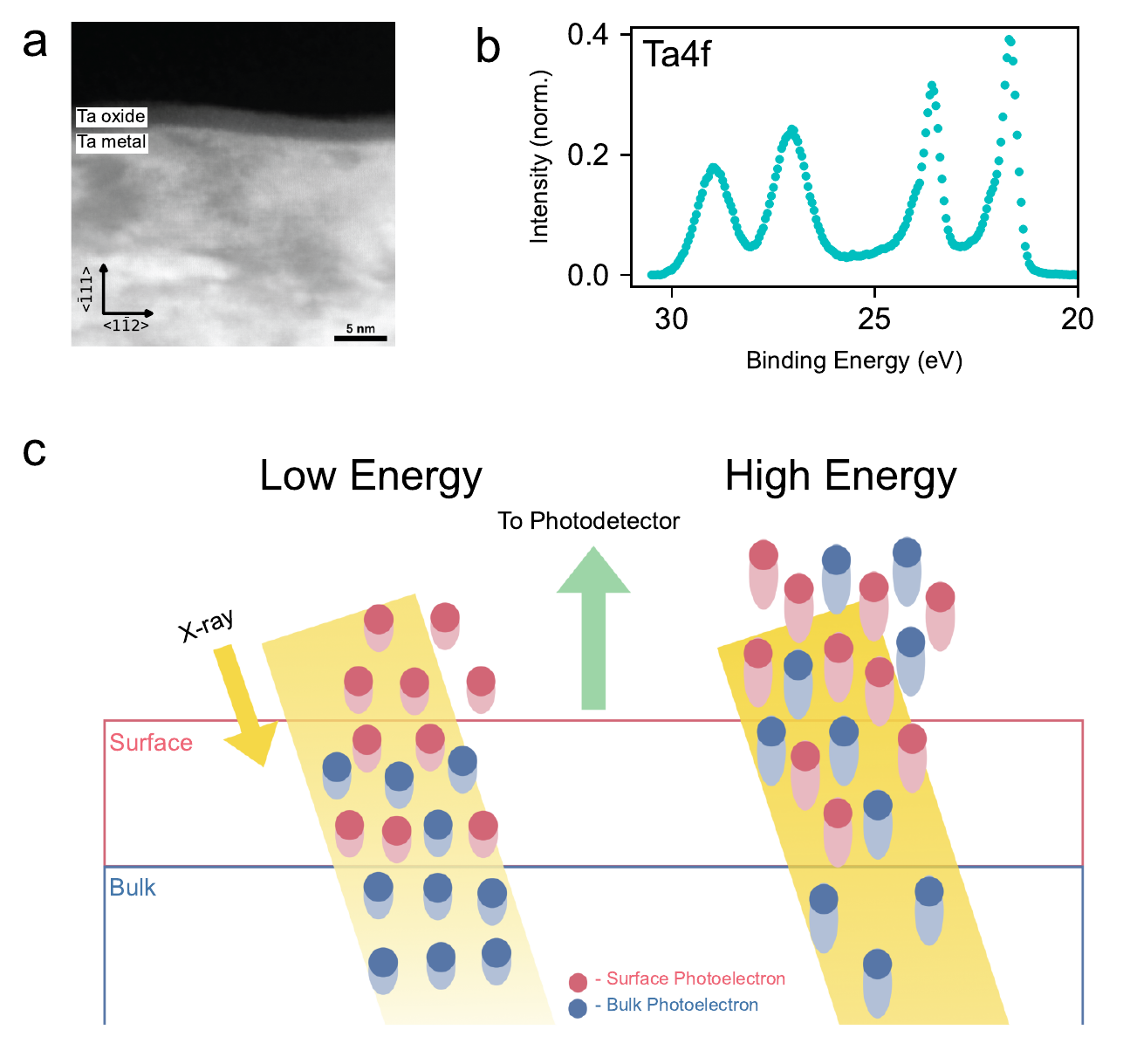}
  \caption{a) High angle annular dark field scanning transmission electron microscope image of the cross-section of a tantalum film on sapphire. The tantalum film has a BCC crystal structure and was grown in the (111) orientation on a c-plane sapphire substrate. An amorphous oxide layer can be seen on top of the tantalum at the tantalum air interface. b) Experimental results of the tantalum binding energy spectrum obtained from X-ray photoelectron spectroscopy (XPS) performed using 760 eV incident photon energy. Each oxidation state of tantalum contributes a pair of peaks to the spectrum due to spin-orbit splitting. At the highest binding energy (26 eV to 30 eV), there is a pair of peaks corresponding to the Ta$^{5+}$ state. At the lowest binding energy, we see a pair of sharp asymmetric peaks corresponding to metallic tantalum (21 eV to 25 eV). c) Schematic explaining the physics behind variable energy X-ray photoelectron spectroscopy (VEXPS). The red and blue dots correspond to photoelectrons excited from a surface oxidation state and bulk oxidation state of the tantalum films respectively. When low energy X-rays are incident on the film surface, photoelectrons are excited with low kinetic energy (depicted by a small tail on the dots). These low energy photoelectrons have a shorter mean free path so that only those emitted from the surface species (colored red) will exit the material and impinge on the detector. When high energy X-rays are incident on the film surface, photoelectrons with high kinetic energy are excited (depicted by a longer tail on the dots). These higher energy photoelectrons have comparatively longer mean free paths so that electrons from the bulk of the film will exit the material alongside electrons from the surface. In our experiment, the angle between the surface and the incident X-rays varies between 6\degree{} and 10\degree{}; the X-rays in this image are shown at a steeper angle for legibility.}
  \label{fig1}
\end{figure}

For all experiments, we use 200 nm thick $\alpha$-Ta films grown on 500 $\mu$m thick c-plane sapphire substrates by DC magnetron sputtering \cite{place_new_2021, kevin_resonator}. A typical tantalum film is covered by a thin layer of amorphous oxide, as shown by its cross-sectional scanning transmission electron microscope (STEM) image (Figure \ref{fig1}a). All samples studied throughout the main text of this paper are from the same deposition, which has a body-centered cubic (bcc) crystal structure ($\alpha$ phase) with a (111) out-of-plane orientation. All samples are cleaned in a 2:1 H$_2$SO$_4$:H$_2$O$_2$ piranha bath for 20 minutes (Section S1 in the Supporting Information).

We use X-ray photoelectron spectroscopy (XPS) to study the oxide at the surface of Ta (Figure \ref{fig1}b). In the 20 eV to 30 eV binding energy range, we observe two prominent pairs of peaks associated with Ta4f core electron ionization. Each oxidation state of Ta appears as a doublet attributed to Ta4f$_{7/2}$ and Ta4f$_{5/2}$ because of significant spin-orbit coupling \cite{Moulder:1992}. The pair of peaks at 22 eV and 24 eV have been previously assigned to metallic Ta$^{0}$, while the pair at 27 eV and 29 eV are assigned to Ta$^{5+}$, here in the form of Ta$_2$O$_5$. The full width half maximum linewidth of the metallic peaks (approximately 0.5 eV) is narrower than those of the oxide (approximately 1.0 eV), likely arising from a higher degree of disorder in the oxide, which is amorphous (Figure \ref{fig1}a). We can also observe shoulder peaks on the higher binding energy side of the Ta$^{0}$ peaks. In addition to these resolvable peaks, there is a broad baseline in the binding energy range between the Ta$^{0}$ and Ta$^{5+}$ peaks, arising from intermediate oxidation states of Ta. In general, a single XPS scan is insufficient to constrain the number of peaks and their associated linewidths, and it does not provide information on how the different oxidation states of tantalum are distributed through the depth of the film.

In order to disentangle the different oxidation states and study their spatial distribution, we perform VEXPS using 17 different incident photon energies in the range of 630 eV to 6000 eV. At lower photon energies, photoelectrons have lower kinetic energy and shorter inelastic mean free paths (IMFP); therefore, lower photon energies are more surface sensitive \cite{Moulder:1992}. The kinetic energy and the IMFP increase with increasing photon energy; thus, the XPS measurements become more bulk sensitive (Figure \ref{fig1}c). We enhance the surface sensitivity of the low incident energy scans by using grazing incidence, which further boosts the depth sampling contrast between photon energies due to the attenuation of X-ray photons through the material. By studying the relative fractions of photoelectrons collected from different oxidation states at different incident photon energies, we may infer the depth distribution of the various tantalum oxidation states.

\section{Results and Discussion}

To quantify the relative abundance of each oxide species, we fit the full dataset of VEXPS spectra with different energies simultaneously. Fitting all spectra at once constrains the relative peak positions, relative peak intensities, peak widths, and skewnesses, which reduces the number of free fit parameters and increases our confidence in the fitted photoelectron intensities. In addition to the Ta$^{5+}$ and Ta$^{0}$ doublets, another three doublets of lower intensity are required to satisfactorily fit the spectra for all photon energies. These include one doublet at approximately 0.4 eV higher binding energy than the Ta$^0$ doublet, which is most prominent at intermediate energies, and two sets of doublets that cannot be individually resolved, located between 21 eV and 22 eV (Ta4f$_{7/2}$) and between 24 eV and 26 eV (Ta4f$_{5/2}$). 

\begin{figure}[h]
\centering
  \includegraphics[width=5 in]{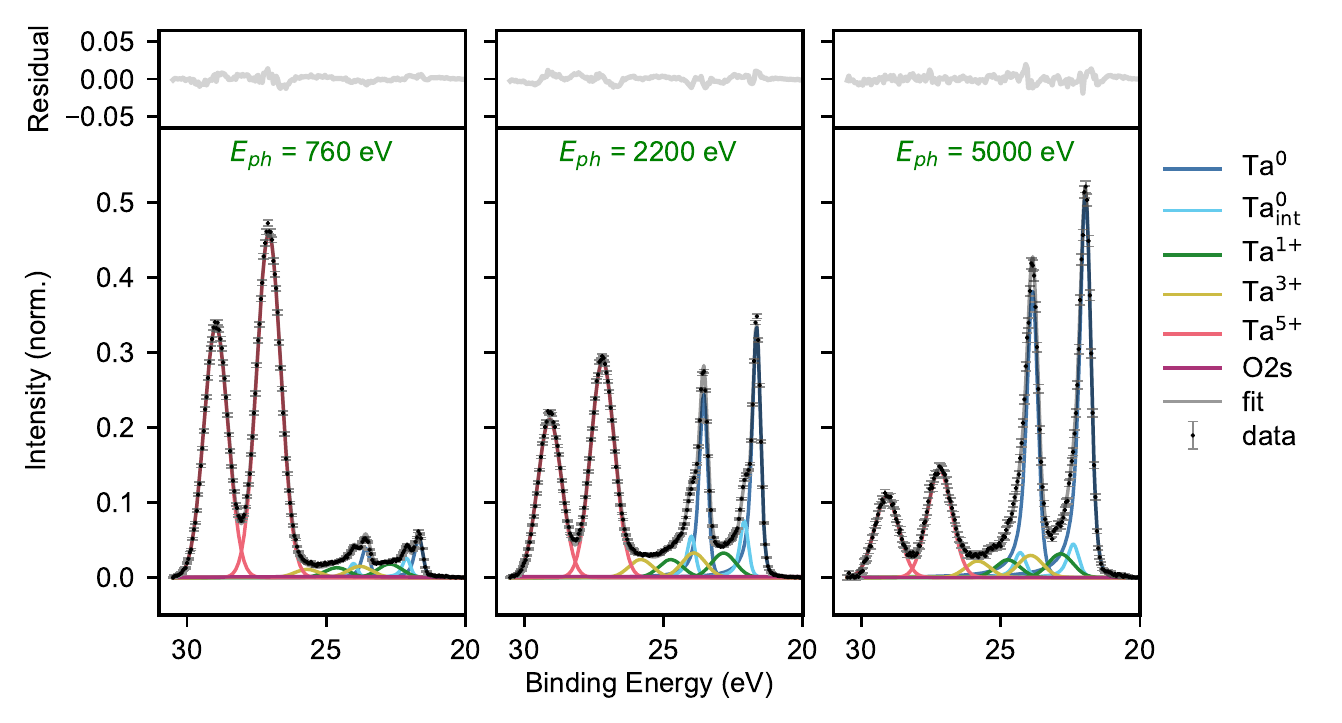}
  %Insert Figure 2 as a 2-column figure
  \caption{Shirley background corrected XPS spectra of Ta4f binding energy obtained at three different incident photon energies. Left panel: with 760 eV X-ray photons, the Ta$^{5+}$ peaks dominate over the Ta$^{0}$ peaks. Middle panel: at 2200 eV photon energy, there is almost equal contribution of photoelectrons from Ta$^{0}$ and Ta$^{5+}$. Right panel: At 5000 eV photon energy, the dominant photoelectron contribution is coming from Ta$^{0}$. In all three plots there is non-zero intensity between the Ta$^{5+}$ and metallic tantalum peaks, indicating minority tantalum oxidation states. The complete set of data and fits corresponding to all 16 incident X-ray energies is shown in Section S3.3 in the Supporting Information. The data are fit with Gaussian profiles for the Ta$^{5+}$, Ta$^{3+}$, and Ta$^{1+}$ species, and skewed Voigt profiles for the Ta$^{0}$ and Ta$^{0}_{\text{int}}$. Included in the fit is also a Gaussian profile corresponding to the O2s peak; the amplitude of this peak is fixed to 5\% of the measured O1s peak intensity.}
  \label{fig2}
\end{figure}

The Ta$^0$ peaks and their shoulder peaks offset by approximately 0.4 eV have similar linewidths, indicating that they both arise from metallic states, so we fit them both with skewed Voigt profiles \cite{sherwood_use_2019}. All other peaks that exhibit much broader linewidths are fit with Gaussian profiles \cite{sherwood_use_2019}. A Shirley background correction is applied to all data prior to XPS peak fitting \cite{engelhard_introductory_2020}, and the data are normalized so that the total intensity under the curve is unity. Lastly, we account for the O2s photoelectron emission that overlaps with Ta4f peaks \cite{Moulder:1992} by measuring the O1s peak. We use the ratio of the photo-ionization cross-sections for the O1s and O2s photoelectrons \cite{O2s_cross_section:2022} to estimate the photoelectron intensity of the O2s peak, neglecting the impact of the kinetic energy difference between O1s and O2s photoelectrons. We find the contribution from the O2s peak to be only a few percent of the total photoelectron intensity for all energies. The background corrected spectra and the results of our peak fitting algorithm are shown in Figure \ref{fig2} for three different photon energies. The Ta$^0$ peak intensities increase with increasing photon energy, while the Ta$^{5+}$ peak intensities decrease, as expected for an oxide layer at the surface. More information on the fitting procedure, as well as the data and fits for all spectra, can be found in Section S3 in the Supporting Information.

We identify the doublet shifted by 0.4 eV from the Ta$^0$ doublet as the layer of tantalum metal closest to the metal-oxide interface, as described in \cite{Himpsel_core_level_shift:1984}. The differing coordination number of tantalum atoms in this layer results in a higher binding energy. We denote this interface species as Ta$^0_{\text{int}}$. We identify the two remaining doublets between 22 eV and 26 eV as Ta$^{1+}$ and Ta$^{3+}$ species based on their locations relative to the Ta$^0$ and Ta$^{5+}$ peaks \cite{Himpsel_core_level_shift:1984}. These additional oxidation states could arise from suboxides of tantalum or other materials, such as tantalum nitride or tantalum carbide. In a wide XPS survey scan we detect no significant atomic species other than tantalum, oxygen, and carbon, thus excluding the presence of nitrides. In a different sample from the same film, we sputtered the top layer with an in-situ argon ion mill, and found that when the C1s peak was removed, the O1s spectrum was largely unchanged, and the Ta$^{1+}$ and Ta$^{3+}$ peaks remained (Section S5 in the Supporting Information). Thus the carbon is present only as adventitious carbon on the top surface of the film, and the Ta$^{1+}$ and Ta$^{3+}$ species arise from suboxides. We therefore assign the Ta$^{1+}$ and Ta$^{3+}$ species as amorphous Ta$_2$O and Ta$_2$O$_3$, respectively.

We obtain a depth dependent chemical profile of the oxide by modelling the dependence of the photoelectron intensity fraction of each species on the incident photon energy. The sample is modeled as a multi-component thin film with five distinct species. Of these five species, the bottom species, Ta$^0$, is assumed to have thickness beyond the sampling depth of our measurements, which we model as being infinite. We assume that the sample composition varies only along the depth $x$, and that it is homogeneous across the plane. The five species are spatially mixed through the depth, described by a set of depth-dependent volume fractions, $\{F_n(x)\}$, with $n$ indexing the 5 different species. The volume fractions are constrained by the sum rule of $\sum_{n=1}^5 F_n(x) = 1$ and a limiting case for the tantalum metal substrate, $\lim_{x \rightarrow\infty} F_{\text{Ta}^0}(x) = 1$.

At each photon energy $E_{ph}$, the intensity fraction, $W_n$, of photoelectrons from species $n$ is obtained by normalizing its photoelectron flux $A_n(E_{ph})$ against the total flux, i.e.,
\begin{equation}\label{eq:W}
W_n(E_{ph}) = \frac{ A_n(E_{ph})}{\sum_{m=1}^5A_m(E_{ph})}.
\end{equation}
The photoelectron flux $A_n$ can be computed for each given set of volume fractions $\{F_n(x)\}$, by considering the attenuation of both the X-rays and the photoelectrons along their respective travel paths:
\begin{equation} \label{eq:A_n model}
    A_n(E_{ph}) = \gamma \eta \rho_n \int_0^\infty F_n(x) 
    \exp \bigg[-\int_0^x \frac{\mu(E_{ph})}{\rho_0} \frac{\rho(s)}{\sin\Theta_{ph}} ds - \frac{x}{\lambda_{el}(E_{el})\sin\Theta_{el}}\bigg] dx,
\end{equation}
where $\gamma$ is the photoelectron collection efficiency; $\eta$ is the photoelectron yield; $\rho_n$ is the density of species $n$; $\rho(s) = \sum\limits_nF_n(s)\rho_n$ is the total density at depth $s$; $\mu/\rho_0$ is the X-ray mass attenuation coefficient; $\Theta_{ph}$ is the angle between the X-ray beam and the sample surface; $\Theta_{el}$ is the angle between the electron detector and the sample surface; $\lambda_{el}$ is the effective electron attenuation length; and $E_{el} = E_{ph}-E_b$ is the kinetic energy of the photoelectron, where $E_b$ is the binding energy of the core level. Derivation of Equation \ref{eq:A_n model} is detailed in Section S4.2 in the Supporting Information. Further discussion can be found in \cite{jablonski_effective_2020}. 

Several parameters in Equation \ref{eq:A_n model} are needed to properly evaluate the intensity fraction $W_n$ in Equation \ref{eq:W}. Our experimental configuration has $\Theta_{el} = 80\degree$ and $\Theta_{ph}=6\degree$ (=$10\degree$) for $E_{ph}<2000$ eV ($\geq2000$ eV). The values of  photoelectron yield $\eta$ and photoelectron collection efficiency $\gamma$ are assumed to be independent of species, so that they cancel in computing $W_n$. The values for $\mu(E_{ph})/\rho_0$ are obtained from those tabulated in \cite{X_ray_attenuation_Henke:1993}. The effective electron attenuation length $\lambda_{el}$ is assumed to be independent of the tantalum species and computed using the empirical relation \cite{EAL_albedo:2020}
\begin{equation}\label{eq:EAL}
\lambda_{el}(E_{el}) = \lambda_{im}(E_{el})[1-0.836\omega(E_{el})],
\end{equation}
as a function of electron kinetic energy $E_{el}$, where $\lambda_{im}$ is the inelastic mean free path of electrons in tantalum and $\omega$ is the single-scattering albedo of tantalum, both tabulated versus the electron kinetic energy \cite{IMFP_tabulated:2015,EAL_albedo:2020}. To further simplify the computation, it is noted that $E_b$ in Equation \ref{eq:A_n model} is well approximated by the average binding energy of all tantalum species (24 eV), because $E_b$ does not vary appreciably and $E_{ph}\gg E_b$. As such, we use $E_{el}\equiv E_{ph}-24$ eV (neglecting work function differences). 

In principle, the volume fractions $\{F_n(x)\}$ can be obtained by fitting $W_n(E_{ph})$ specified by Equation \ref{eq:W} and Equation \ref{eq:A_n model} to the experimental photoelectron intensity fractions. In practice, a parameterization of $\{F_n(x)\}$ is needed to limit the number of independent fitting parameters and to avoid overfitting. $\{F_n(x)\}$ are parameterized using a basis set of smooth, analytic functions that are normalized by themselves, i.e., $\sum_{n=1}^5 F_n(x) = 1$ for all depth $x$ (Section S4.3 in the Supporting Information).

\begin{figure}[h]
\centering
  \includegraphics[width=5 in]{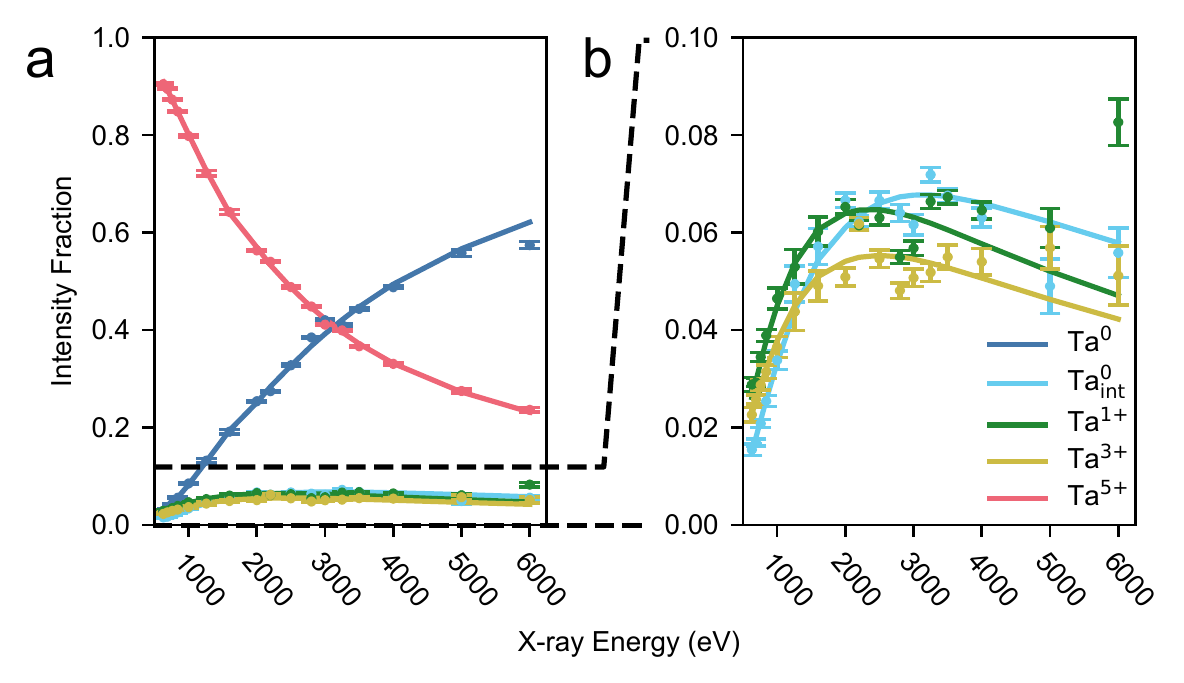}
  %Insert Figure 3 as a 2-column figure
  \caption{a) Relative photoelectron fraction for the different tantalum oxidation states as a function of incident X-ray photon energy. The dots are experimental data extracted from fitting Ta4f spectra at different incident X-ray energies. Solid lines represent an iterative fit to the data using Equation \ref{eq:W} and Equation \ref{eq:A_n model}, corresponding to the expected intensity fractions from the depth profile shown in Figure \ref{fig4}a. b) Close up view of relative photoelectron fraction contribution from the Ta$^{3+}$, Ta$^{1+}$, and Ta$^{0}_{\text{int}}$. The rise, plateau, and fall of the photoelectron intensity fractions with X-ray energy indicate that these three species are localized at a buried interface.}
  \label{fig3}
\end{figure}

\begin{figure}[h]
\centering
  \includegraphics[width=3.375 in]{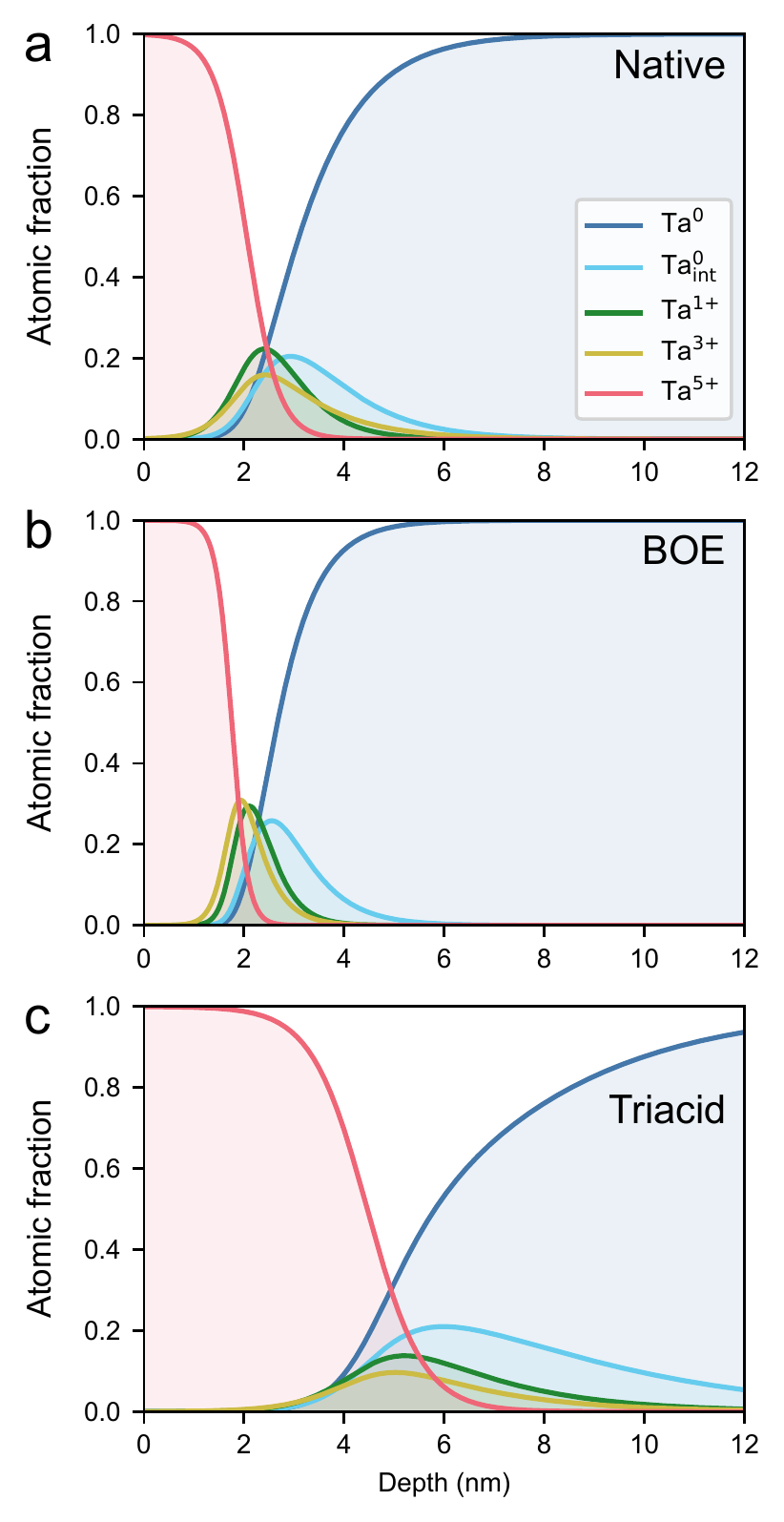}
  %Insert Figure 4 as a 1-column figure
  \caption{Extracted chemical profiles for three different samples: a) native b) BOE treated and c) triacid treated. Fitted photoelectron intensity fractions are shown in Section S4.6 in the Supporting Information.}
  \label{fig4}
\end{figure}

For a film that has only undergone piranha cleaning (``native") film, the photoelectron intensity fraction of Ta$^0$ increases monotonically and that of Ta$^{5+}$ decreases monotonically with photon energy (Figure \ref{fig3}a). The photoelectron intensity fractions of Ta$^{1+}$, Ta$^{3+}$, and Ta$^0_{\text{int}}$ initially increase with photon energy, but then peak and slowly decrease (Figure \ref{fig3}b). These observations indicate that the Ta$^0$ species is located in the bulk, the Ta$^{5+}$ species is located at the surface, and the other species are located at an intermediate depth. The resulting depth profile shows a surface layer of Ta$^{5+}$ which extends approximately 2 nm into the depth, an approximately 1 nm to 2 nm thick region with overlapping Ta$^{3+}$, Ta$^{1+}$, and Ta$^{0}_{\text{int}}$ profiles, and a bulk layer of Ta$^0$ (Figure \ref{fig4}a).

Quantitative depth profiling allows us to make detailed comparisons between different films. In addition to measuring the native tantalum oxide after our sputter deposition process, we also measure the tantalum oxide after two different surface treatments. The first is immersing the film for 20 minutes in 10:1 buffered oxide etch (BOE) to etch the oxide, and the second is refluxing the film in 1:1:1 nitric, perchloric, and sulfuric acids for two hours (``triacid") to grow the oxide (Section S2 of Supporting Information). The results of the depth profile analysis for each case are shown in Figure \ref{fig4}b-c with the fitted photoelectron intensities shown in Section S4.6 in the Supporting Information. We also extract an effective thickness of each tantalum oxide species and the tantalum interface species by integrating the area under each curve. The effective thicknesses for the native, BOE treated, and triacid treated films are tabulated in Table \ref{table1}.

The BOE treated film exhibits statistically significantly smaller effective thicknesses than the native film for the Ta$^{5+}$, Ta$^{3+}$, and Ta$^{1+}$ species (approximately 20$\%$ lower in all cases), while the Ta$^0_{\text{int}}$ layer effective thickness does not show a significant difference (Table \ref{table1} and Figure \ref{fig4}). The triacid treated film shows significantly larger thicknesses for all four species compared to either the native or BOE treated film. Furthermore, the BOE treated film has a narrower distribution of the Ta$^{1+}$, Ta$^{3+}$, and Ta$^0_{\text{int}}$ species compared to the native film while the triacid film has significantly broader distributions of those three species.

Based on our measured etch rate, the BOE treatment does not etch away the entire Ta$^{5+}$ layer (Section S6 in the Supporting Information), and the mechanism for BOE interaction with the buried interface is unclear. Using atomic force microscopy (AFM) on the film profiled in Figure \ref{fig4}a, we observe a root mean square surface roughness of 0.568 nm with a minimum observed depth of 1.8 nm across a 500 nm square area (Supporting Information Section S7). These two values are a significant fraction of the 2.257 nm $\pm$ 0.023 nm thick Ta$^{5+}$ layer we found for our native oxide film. We have also observed small pinholes of similar depth variation in samples with smoother morphology, which exhibit the same chemical profile in VEXPS (Section S7 in the Supporting Information). We hypothesize that surface roughness and pinholes allow access to the buried interface. The X-ray spot size in our VEXPS experiments was an ellipse with major and minor diameters approximately 300 $\mu$m and 50 $\mu$m, which is much larger than the surface roughness features we observe. Therefore the VEXPS measurements reflect an average over many of the surface roughness features.

We can use VEXPS data to elucidate potential sources of microwave loss. In \cite{kevin_resonator}, using data from native, BOE treated, and triacid treated samples, we estimated that if the microwave dielectric loss tangent of the tantalum oxide scales with the thickness of the oxide layer, then contributions to the microwave dielectric loss tangents from the tantalum oxide and adventitious carbon species are comparable. However, the chemical profiles of the native, BOE treated, and triacid treated samples show that the thicknesses and distributions of each of the Ta$^{5+}$, Ta$^{3+}$, Ta$^{1+}$, and Ta$^{0}_\text{int}$ species vary with surface treatments. Therefore changes in the dielectric loss tangent could arise from changes to any combination of the interface species. For example, based on these detailed chemical profiles, an alternative plausible hypothesis for the observations in \cite{kevin_resonator} would be that the dielectric loss arises entirely from the Ta$^{3+}$ layer rather than separate contributions of the oxide and adventitious hydrocarbons. Independently varying the thickness and distribution of the Ta$^{5+}$, Ta$^{3+}$, Ta$^{1+}$, and Ta$^{0}_\text{int}$ over many more surface compositions than those measured in \cite{kevin_resonator} could be used to determine a precise, atomistic model for dielectric loss.

In addition to comparing depth profiles across differently treated tantalum films, we can compare the depth profile of the native tantalum film (Figure \ref{fig4}a) to a similar depth profile from a niobium surface presented in \cite{Anjali_Nb:2021}, in which depth profiles of various niobium films were measured and compared to coherence times of qubits measured on each film. The total oxide layer thickness is appreciably smaller in tantalum versus niobium, and the tantalum suboxide species are more confined to the oxide-metal interface. These differences suggest that the comparatively thinner oxide and more confined suboxide layer of tantalum may be one reason why qubits made out of tantalum films show longer coherence times than those made out of niobium films. This observation is consistent with recent measurements of niobium resonators with thinner oxide and correspondingly higher quality factors \cite{verjauw_investigation_2021}.

\section{Conclusion}

Variable energy X-ray photoelectron spectroscopy and chemical profiling reveals that in addition to the dominant Ta$^{5+}$ oxide and previously known Ta$^{0}_{\text{int}}$ species, there also exist two tantalum suboxide species that we have identified as Ta$^{1+}$ and Ta$^{3+}$. These two suboxides are localized in depth at the interface between the Ta$^{5+}$ layer and the bulk Ta$^{0}$. We observe that the tantalum metal-air interface contains a smaller fraction of minority species and has a thinner majority oxide species than a corresponding Nb interface \cite{Anjali_Nb:2021}. Our measurements on BOE treated and triacid treated films show that the tantalum oxide is surprisingly robust, but can be altered slightly. Correlating these measurements with systematic measurements of quantum devices based on Ta films using many more surface treatments would elucidate the quantitative contributions of various oxide and interface species to dielectric loss, guiding future work on device optimization. More broadly, our method of data collection and analysis can provide a foundation for future studies of the interface layers of tantalum thin films or thin films of other metals to understand the structure of those interfaces and effects of surface treatments.

\medskip
\textbf{Supporting Information} \par 
Supporting Information is available from the Wiley Online Library or from the author.

% Acknowledgements
\medskip
\textbf{Acknowledgements} \par %delete if not applicable))
The authors acknowledge Mayer Feldman for support with tantalum depositions and Esha Umbarkar for sample preparation. We also acknowledge Denis Potapenko for support with the XPS system in the Princeton Imaging and Analysis Center.

This material is based upon work primarily supported by the U.S. Department of Energy, Office of Science, National Quantum Information Science Research Centers, Co-design Center for Quantum Advantage (C$^2$QA) under contract number DE-SC0012704. Film characterization and processing was partly supported by the National Science Foundation (RAISE DMR-1839199). This research used resources of the Spectroscopy Soft and Tender Beamlines (SST-1 and SST-2) of the National Synchrotron Light Source II and the Electron Microscopy and Materials Synthesis \& Characterization facilities of the Center for Functional Nanomaterials (CFN), U.S. Department of Energy Office of Science Facilities at Brookhaven National Laboratory under contract no. DE-SC0012704. The authors acknowledge the use of Princeton’s Imaging and Analysis Center (IAC), which is partially supported by the Princeton Center for Complex Materials (PCCM), a National Science Foundation (NSF) Materials Research Science and Engineering Center (MRSEC; DMR-2011750). Some chemical treatments were performed in the Princeton Institute for the Science and Technology of Materials (PRISM) cleanroom. This project was supported in part by the U.S. Department of Energy, Office of Science, Office of Workforce Development for Teachers and Scientists (WDTS) under the Science Undergraduate Laboratory Internships Program (SULI).

\begin{table}[h]
 \caption{Effective thickness of different tantalum oxidation states as obtained from depth profile fitting for different tantalum films. All data in nm. Uncertainties are $\pm{}1\sigma{}$ confidence intervals reflecting uncertainty in our fit.}
 \label{table1}
 \centering
 \setlength{\tabcolsep}{10pt}
  \begin{tabular}[htbp]{@{}lllll@{}}
    \hline
    Film & Ta$^{5+}$ & Ta$^{3+}$ & Ta$^{1+}$ & Ta$^{0}_{\text{int}}$ \\
    \hline
    Native & 2.257 $\pm$ 0.023 & 0.370 $\pm$ 0.016 & 0.370 $\pm$ 0.017 & 0.368 $\pm$ 0.019\\
    Ta treated in 10:1 BOE &  1.853 $\pm$ 0.028 & 0.296 $\pm$ 0.022 & 0.302 $\pm$ 0.023 & 0.400 $\pm$ 0.021\\
    Ta treated in triacid & 4.826 $\pm$ 0.036 & 0.379 $\pm$ 0.016 & 0.545 $\pm$ 0.020 & 1.198 $\pm$ 0.027\\
    \hline
    
  \end{tabular}
\end{table}

\pagebreak

\bibliographystyle{MSP}
%\bibliography{MSP-template}
\bibliography{references}

\end{document}

\pagestyle{fancy}

\title{Supporting Information for: Chemical profiles of the oxides on tantalum in state of the art superconducting circuits}

\maketitle

\author{Russell A. McLellan$^\dag{}$}
\author{Aveek Dutta$^\dag{}$}
\author{Chenyu Zhou}
\author{Yichen Jia}
\author{Conan Weiland}
\author{Xin Gui}
\author{Alexander P. M. Place}
\author{Kevin D. Crowley}
\author{Xuan Hoang Le}
\author{Trisha Madhavan}
\author{Youqi Gang}
\author{Lukas Baker}
\author{Ashley R. Head}
\author{Iradwikanari Waluyo}
\author{Ruoshui Li}
\author{Kim Kisslinger}
\author{Adrian Hunt}
\author{Ignace Jarrige}
\author{Stephen A. Lyon}
\author{Andi M. Barbour}
\author{Robert J. Cava}
\author{Andrew A. Houck}
\author{Steven L. Hulbert}
\author{Mingzhao Liu*}
\author{Andrew L. Walter*}
\author{Nathalie P. de Leon*}

$^\dag{}$These authors contributed equally.

\begin{affiliations}

\end{affiliations}

\setlength{\parindent}{20pt}

\section{Materials}

All samples used in the variable energy X-ray photoelectron spectroscopy (VEXPS) measurements were approximately 200 nm thick $\alpha$-Ta(111). Measurements reported in the main text were performed on a film deposited by DC magnetron sputtering onto c-plane sapphire by Star Cryoelectronics.  Measurements reported in Section \ref{Effect of multiple BOE treatments} were performed on a film deposited by DC magnetron sputtering onto c-plane sapphire at Princeton University. Phase and orientation of both tantalum films were confirmed by X-ray diffraction and measurements of both the superconducting critical temperature and critical magnetic field in a Physical Property Measurement System. All samples were approximately 7 mm x 4 mm rectangles. Morphology differences between the two films are described in Section \ref{AFM_section}.

\section{Methods} \label{sec:methods}

10:1 buffered oxide etch (BOE) is a mixture of 10 parts 40\% NH$_4$F solution to 1 part 49\% HF solution by volume. We procured 10:1 BOE from Transene. BOE treated samples were placed in buffered oxide etch at room temperature and were not agitated. After 20 minutes, the samples were removed and triple rinsed in de-ionized water and 2-propanol before being blown dry in N$_2$.

The triacid treatment is 1:1:1 equal mix by volume of 95-98\% H$_2$SO$_4$, 70\% HNO$_3$, and 70\% HClO$_4$ solutions (all percentages by weight). We procured all solutions from SigmaAldrich (catalogue numbers: H$_2$SO$_4$ - 258105, HNO$_3$ - 225711, HClO$_4$ - 244252). After the sample was added to the mixture, it was heated to 200 $\degree$C for 2 hours and then allowed to cool for 1 hour. During this process, the exhaust gas was cooled and bubbled through water. No agitation was performed. After cooling, the sample was removed, triply rinsed in de-ionized water and 2-propanol before being blown dry in N$_2$.

All films were treated in piranha solution for 20 minutes. BOE and triacid treated samples were treated in piranha solution prior to undergoing BOE or triacid treatments. Native samples were treated several hours before being inserted into the vacuum chamber for VEXPS. Piranha solution was prepared with 2 parts H$_2$SO$_4$ to 1 part H$_2$O$_2$ by volume, initially at room temperature. No external heating or agitation was performed. After being removed from the piranha solution, samples were triply rinsed in de-ionized water and 2-propanol before being blown dry in N$_2$. The effect of the piranha treatment is explored in Section \ref{sec:piranha}.

VEXPS measurements were performed at the Spectroscopy Soft and Tender-1 and Spectroscopy Soft and Tender-2 (SST-1 and SST-2) beam lines at the National Synchrotron Light Source II at Brookhaven National Laboratory. SST-1 was used for X-ray energies less than 2000 eV and SST-2 was used for X-ray energies greater than or equal to 2000 eV. The difference between SST-1 and SST-2 is in the energy range of the X-ray beam sent to the sample; beam lines share the same vacuum chamber. The step size for VEXPS measurements was 0.05 eV and dwell time was 100 ms. The detector pass energy was varied from 20 eV to 200 eV as X-ray energy was changed based on the observed electron counts and whether we could resolve the Ta$^0_\text{int}$ shoulder peak. Depending on the experiment, a separate sample of either silver or gold was scanned at each X-ray energy as a binding energy reference. When a silver reference was used, we set the binding energy of the Ag3d$_{5/2}$ peak to 368.3 eV. When a gold reference was used, we set the binding energy of the Au4f$_{7/2}$ peak to 84 eV.

\section{XPS trace analysis} \label{XPS_trace_analysis}
\subsection{Uncertainty calibration of XPS data}
The number of electron counts varies significantly from datapoint to datapoint. The maximal number of counts for different spectra can vary by over two orders of magnitude across the X-ray energy we scanned. This difference is largely attributed to differences in incident X-ray photon flux. We expect the uncertainties in our measured photoelectron intensity to be a function of the number of counts, and we need to  calibrate the uncertainties to ensure that we are fitting the peaks correctly.

To calibrate our error bars, we fit a line to regions of traces that contain no peaks. The Ta4f peaks are close in binding energy to the Ta5p and Ta5s peaks, and therefore it was not possible to find a region near the Ta4f peaks that contained only background counts. However, each time we measured Ta4f traces at a particular X-ray energy, we also measured the O1s and C1s peaks on each tantalum sample. We also have Ag3d spectra on the reference sample that we used to calibrate the binding energy at each photon energy. We set our binding energy range on these peaks large enough to capture a region several eV wide with no observable satellite loss peaks. As there are no observable peaks in these regions, we assume that a line is the best fit to each dataset.

\begin{figure}
  \centering
  \includegraphics[width=0.75\linewidth]{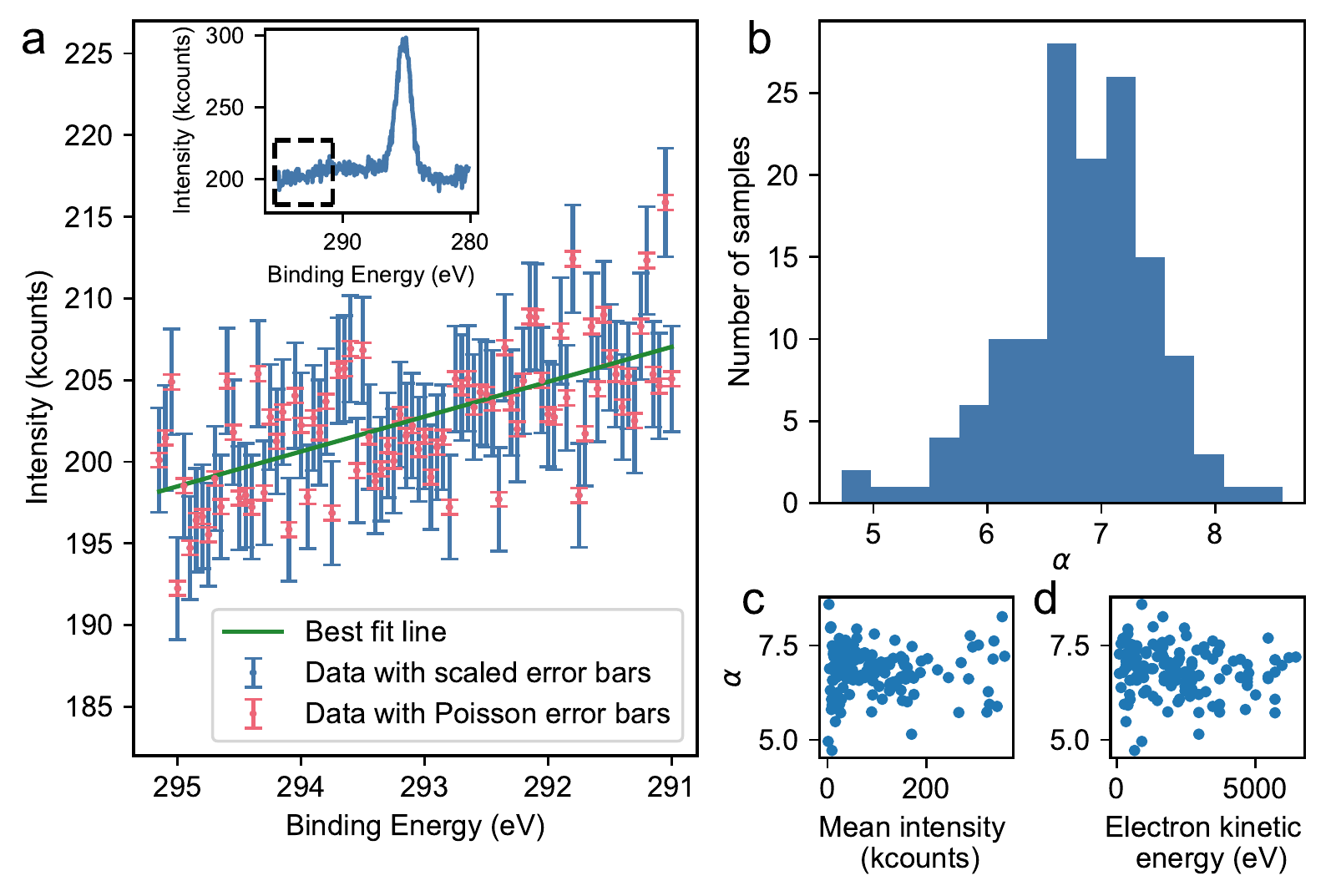}
  \caption{a) Photoelectron counts for a C1s scan between 291 eV and 295 eV at an incident X-ray energy of 2000 eV. We observe no loss peaks in this region. Data is plotted in red with errors bars assuming a Poisson distribution. However, the best fit line to the data (green) does not give a reduced $\chi^2$ of unity with Poisson error bars. The errors bars must be scaled by a number $\alpha$ to give $\chi^2=1$. These scaled error bars are plotted in blue. (Inset: The full C1s scan with the area in the main plot indicated) b) Histogram of the scaling parameter $\alpha$ obtained by scaling error bars in empty regions of all C1s, O1s, and Ag3d regions as in a). The mean value of $\alpha$ is $\overline{\alpha}$ = 6.4 $\pm$ 0.9. c-d) the data in b) plotted versus the mean intensity (c) and electron kinetic energy (d) of each scan. There is no systematic trend with either variable, indicating that a single scaling factor $\overline{\alpha}$ is sufficient to capture the uncertainty in the intensity in all scans.}\label{fig:ErrorScaling}
\end{figure}

We initially assume that the electron count statistics are Poissonian, and therefore $\sigma_{I_i} = \sqrt{I_i}$, where $I_i$ is the number of counts for the $i$th datapoint and $\sigma_x$ is the uncertainty in the measurement $x$. For each C1s, O1s, and Ag3d trace, we fit a line to background regions using these Poissonian error bars, and then scale the uncertainties so that the reduced $\chi^2$ value of the fit is unity. The uncertainties are now given by $\sigma_{I_i} = \alpha{}\sqrt{I_i}$, where $\alpha$ is a scalar. The binding energy regions we considered as the background are shown in Table \ref{table_uncertainty}. The uncertainty scaling is shown for an example C1s trace in Figure \ref{fig:ErrorScaling}(a).

A value of $\alpha$ was fit individually to each O1s, C1s, and Ag3d trace. There is no systematic trend with the value of $\alpha$ versus either the mean electron kinetic energy of the scan or the mean number of electron counts for the scan (Figure \ref{fig:ErrorScaling}(c-d)). All values of $\alpha$ appear to be drawn from a unimodal distribution with mean $\overline{\alpha} = 6.4\pm 0.9$ (Figure \ref{fig:ErrorScaling}(b)). Based on these results, we use error bars $\sigma_{I_i} = \overline{\alpha}\sqrt{I_i}$ for all of our data.

We interpret the value of $\overline{\alpha}$ being larger than unity as consistent with the large amount of gain in the electron detection system. We also note that any overall scaling factor $\overline{\alpha}$ would not affect the fits of the XPS peaks, but only the reduced $\chi^2$ value. The parameters resulting from our fits would be different only if the uncertainties scaled in a manner other than $\sigma_{I_i} \propto \sqrt{I_i}$.

\begin{table}[h]
 \caption{Binding energy regions considered empty for uncertainty calibration.}
 \label{table_uncertainty}
 \centering
 \setlength{\tabcolsep}{10pt}
  \begin{tabular}[htbp]{@{}lccc@{}}
    \hline
     & O1s & C1s & Ag3d   \\
    \hline
    Lower bound (eV) & 538 & 291 & 360 \\
    Upper bound (eV) & 543 & 295 & 363 \\
    \hline
    
  \end{tabular}
\end{table}

\subsection{Background subtraction}

We subtracted a Shirley background from all Ta4f and O1s XPS spectra before fitting. The Shirley background attempts to correct for the step change in background signal seen before and after a large peak, attributed to electrons from the peak which are scattered before being detected at a lower kinetic energy. The initial presentation of the Shirley background is given in \cite{shirley_high-resolution_1972} and it is discussed in more detail in \cite{engelhard_introductory_2020}. The Shirley background correction process is presented here based on those treatments, and we include a method of propagating error through the background correction.

First, we choose binding energies above and below the peaks of interest where we assume that any signal present is only background signal. The intensity of electrons at these binding energies will be used in our background subtraction, so we average over a small number of binding energy points to extract a mean and uncertainty of the background counts.

We subtract a flat background, $\overline{I_1}$, from the spectrum, equal to the mean number of counts at the lower binding energy. This initial flat background correction compensates for any constant source of noise that is independent of the electrons from the peaks. The intensity after the flat background correction is given by:

\begin{equation} \label{eq:flat_background}
    I^{'}_i = I_i - \overline{I_1},
\end{equation}

where $I^{'}_i$ is the flat-background corrected intensity at index $i$, and $I_i$ is the uncorrected intensity at binding energy index $i$. The uncertainty from the flat background is propagated to the uncertainty in the flat-background corrected intensities by:

\begin{equation}\label{eq:flat_error}
    \sigma_{I^{'}_i}^2 = \sigma^2_{I_i} + \sigma^2_{I_1},
\end{equation}
 
 where $\sigma_A$ indicates the uncertainty in the variable $A$.

After the flat background correction, we apply the Shirley background correction. The Shirley background is given by:

\begin{equation} \label{eq:shirley_background}
    s_i = \left(\frac{\sum\limits_{j\leq{}i} I^{'}_j}{\sum\limits_k{I^{'}_k}}\right)(I^{'}_f - I^{'}_1),
\end{equation}

where $s_i$ is the Shirley background at index $i$, and the indices run from 1 to $f$, where $f$ is the index corresponding to the highest binding energy. The Shirley background corrected intensity is given by $I_i^{''}$ = $I_i^{'} - s_i$. To calculate the uncertainty in the Shirley-corrected data, we use the following formula for error propagation \cite{bevington_error_2003}:

\begin{equation} \label{eq:error_propagation}
    \sigma_B^2 = \sum\limits_i \sigma_{b_i}^2 \left( \frac{\partial{}B}{\partial{}b_i} \right)^2,
\end{equation}

where $B$ is a function of the $b_i$s and each $b_i$ has a known uncertainty. We have assumed that all $\sigma_{b_i}$ are independent in Equation \ref{eq:error_propagation}. Applying Equation \ref{eq:error_propagation} to $I^{''}_i$, we find:

\begin{equation}
    \sigma_{I^{''}_i}^2 = \sigma_{I^{'}_i}^2 + g_i^2(I^{'}_f-I^{'}_1)^2 \left(\frac{\sigma_{g_i}^2}{g_i^2} + \frac{\sigma_{I^{'}_f}^2 + \sigma_{I^{'}_1}^2}{(I^{'}_f + I^{'}_1)^2}\right), 
\end{equation}

where:
\begin{align}
    g_i & = \frac{\sum\limits_{j\leq{}i} I^{'}_j}{\sum\limits_k{I^{'}_k}} \\
    \sigma_{g_i}^2 & = \left( \frac{1-g_i}{\sum\limits_k I^{'}_k} \right)^2 \sum\limits_{j\leq{}i} \sigma_{I^{'}_j}^2 + \left( \frac{g_i}{\sum\limits_k I^{'}_k} \right)^2 \sum\limits_{j>i} \sigma_{I^{'}_j}^2.
\end{align}

\begin{figure}
\centering
  \includegraphics[width=0.6\linewidth]{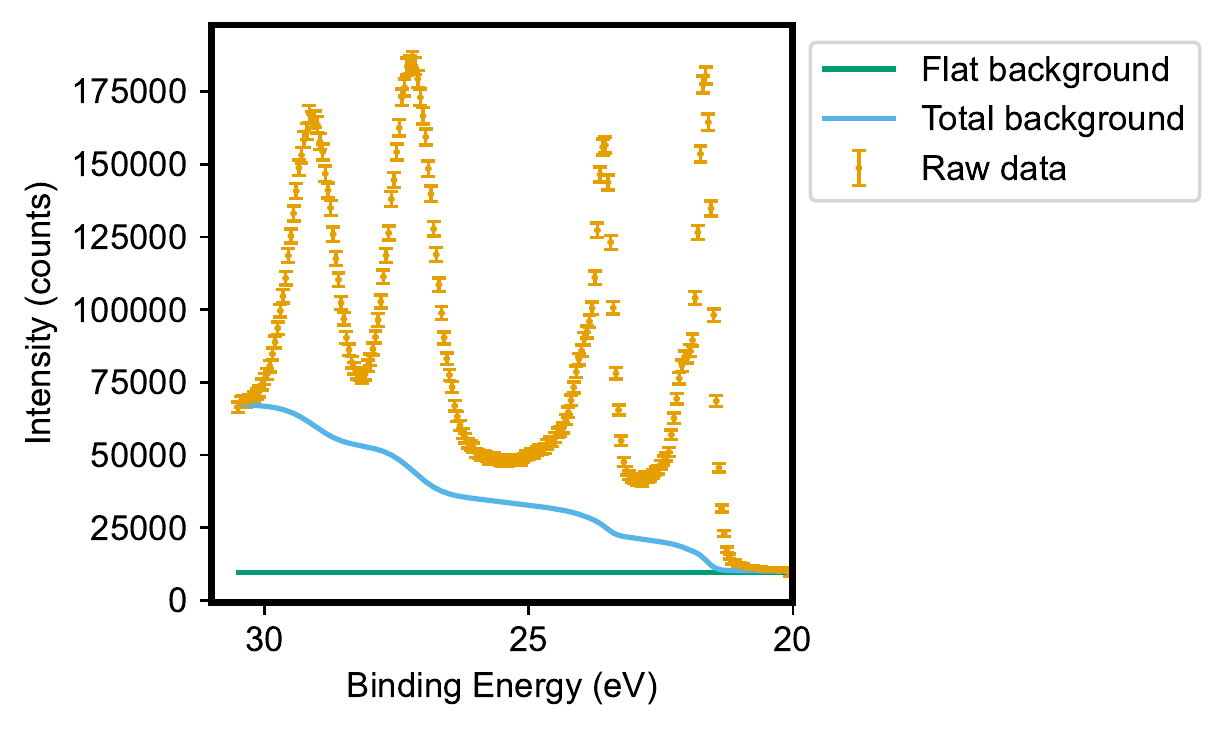}
  \caption{Data and calculated background for the Ta4f spectra measured on the untreated (''Native") sample with an X-ray energy of 2200 eV. Both flat and total (flat + Shirley) backgrounds are shown. The flat background corrects for any constant noise unrelated to the electrons from the peaks, while the Shirley background corrects for electrons from the peaks which are scattered before being collected.}
  \label{fig:sup_background}
\end{figure}

We perform background subtraction for both Ta4f and O1s peaks. For Ta4f peaks, the binding energy range for background subtraction is 20 eV to 30.5 eV. For O1s peaks, the binding energy range for background subtraction is 528 eV to 536 eV. An example of background subtraction on a Ta4f spectrum is shown in Figure \ref{fig:sup_background}.

\subsection{XPS peak fitting with constraints} 

As described in the main text, XPS peaks are fit with either Gaussian (Ta$^{5+}$, Ta$^{3+}$, Ta$^{1+}$, and O2s), or skewed Voigt (Ta$^0$ and Ta$^0_{\text{int}}$) profiles \cite{engelhard_introductory_2020}. Each Ta peak is doubled due to the strong spin orbit coupling of tantalum, leaving us with a total of 11 peaks per Ta4f spectrum. In order to fit these peaks with minimal uncertainty, we fit all Ta4f spectra for each sample simultaneously, constraining peak locations, amplitudes, widths, and skewnesses as described in this section. Our peak fitting is implemented with the lmfit Python package \cite{noauthor_getting_nodate}.

A Gaussian profile is described by:
\begin{equation}\label{eq:gaussian}
    f(x;A,\mu,\sigma) = \frac{A}{\sigma{}\sqrt{2\pi{}}}e^{-(x-\mu{})^2/2\sigma^2},
\end{equation}

where $A$ is the area under the curve, $\mu$ is the center of the profile, and $\sigma$ is the width of the profile. A skewed Voigt profile is described by:
\begin{equation}\label{eq:voigt}
    f(x;A,\mu,\sigma,\lambda) = \frac{A \text{Re}[w(z)]}{\sigma{}\sqrt{2\pi{}}}\left(1+\text{erf}\left[\frac{\lambda(x-\mu)}{\sigma\sqrt{2}} \right] \right),
\end{equation}

where $A$, $\mu$, and $\sigma$ have the same meanings as in Equation \ref{eq:gaussian}, $\lambda$ is the dimensionless skewness parameter, erf is the error function, and $z$ and $w(z)$ are given by:

\begin{align}
    z & = \frac{x-\mu-i\sigma}{\sigma\sqrt{2}} \\
    w(z) & = e^{-z^2}\text{erfc}(-iz). \label{eq:w(z)}
\end{align}

In Equation \ref{eq:w(z)}, erfc($-iz$) = 1-erf($-iz$) is the complementary error function.

Physically, for each peak, $A$ represents the total photoelectron intensity, $\mu$ represents the binding energy of the peak, $\sigma$ characterizes broadening (including broadening introduced by the photoelectron detector), and $\lambda$ characterizes the shake-up satellite structure of conductive compounds. We use these physical interpretations to constrain parameters as follows.

First, we assume that the shake-up satellite structure of the Ta$^{0}_\text{int}$ and Ta$^{0}$ states are identical and unchanged across experiments at different X-ray energies. This assumption implies that a single value of $\lambda$ can be used for all Ta$^{0}_\text{int}$ and Ta$^{0}$ peaks across all X-ray energies for each sample.

Second, each pair of peaks that arise from the Ta$_{7/2}$ and Ta$_{5/2}$ spin states of the same tantalum oxidation state should share the same $\sigma$ parameter. Further, we assume that the proportion of electron population in the Ta$_{7/2}$ and Ta$_{5/2}$ states, the proportion of the X-ray cross sections of the Ta$_{7/2}$ and Ta$_{5/2}$ states, and spin-orbit coupling strengths are independent of the tantalum oxidation state and incident X-ray energy. These assumptions imply that the ratio of $A$ and the difference in $\mu$ between any pair of Ta$_{7/2}$ and Ta$_{5/2}$ peaks is the same

Third, we assume that the binding energy of each peak does not change between experiment at different X-ray energies. In practice, variations on the order of 0.1 eV are observed in the positions of the Ta$^{5+}$ and Ta$^{0}$ peaks, which we attribute to charging of the tantalum oxide layer. Instead of fixing binding energy positions absolutely, we constrain relative peak position. We allow the binding energy of the Ta$^{5+}$ and Ta$^{0}$ peaks to vary at each X-ray energy, as these peaks are easily located. The positions of the Ta$^{1+}$ and Ta$^{3+}$ are constrained relative to the position of the Ta$^{5+}$ peak across all X-ray energies, and likewise the position of the Ta$^{0}_\text{int}$ peak is constrained relative to the Ta$^{0}$ peak. We constrain the O2s peak to have a single binding energy, although in practice, this peak is small and found to be quite wide, so variations in the O2s binding energy are unlikely to have a significant impact on the fit.

Fourth, we assume that for a given XPS spectrum, the broadening of the Ta$^0$ and Ta$^0_\text{int}$ peaks are the same, as is the broadening of the Ta$^{1+}$, Ta$^{3+}$, and Ta$^{5+}$ peaks. We are assuming that only three different values of $\sigma$ are needed for each Ta4f spectrum; one for the tantalum oxide species, one for the metallic species, and one for the O2s peak. In practice, relaxing this assumption does not significantly affect the peak fits.

Fifth, we constrain the amplitude of the O2s peak to 5\% of the amplitude of the O1s spectrum. The X-ray cross-section for the O1s and O2s states are approximately constant in the X-ray energy range 600 eV to 1500 eV \cite{O2s_cross_section:2022}, and we extrapolate this ratio out to our maximum X-ray energy of 7000 eV. The intensity of the O2s peak is typically less than 5\% of the Ta4f intensity, so we do not believe that this extrapolation introduces significant error. Note that we are neglecting the kinetic energy difference between an O1s photoelectron and an O2s photoelectron. At each X-ray energy, we numerically integrate the area under a background corrected O1s spectrum between binding energies 528 eV and 536 eV to calculate the O1s intensity, and this value is scaled to fix the O2s energy at the corresponding Ta4f spectrum. 

\begin{figure}
\centering
  \includegraphics[width=\linewidth]{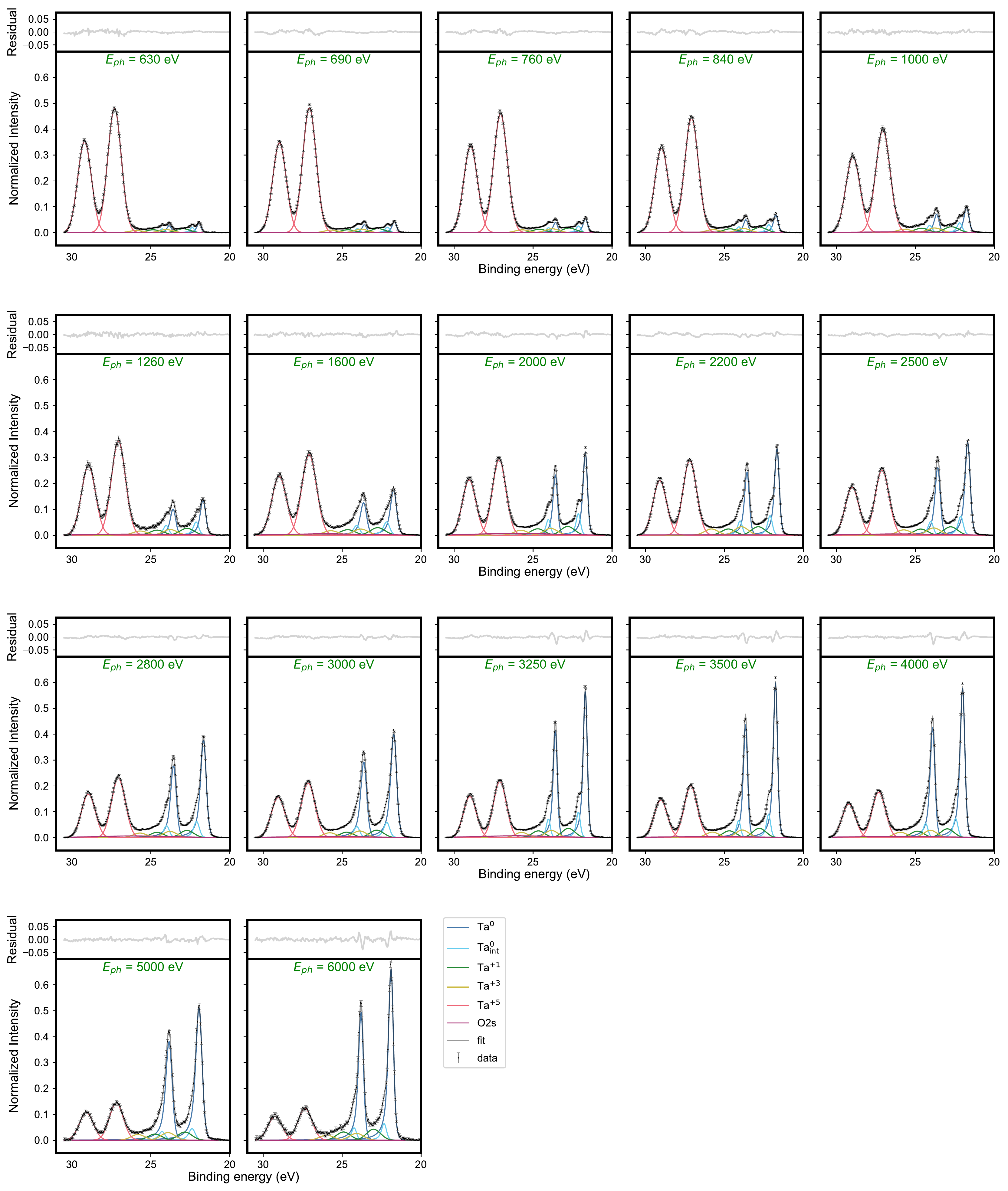}
  \caption{Fitted Ta4f intensity spectra for all X-ray energies on the untreated (''Native") sample. All spectra are fitted simultaneously with certain parameters constrained between spectra, as described in the text. Three of these fitted spectra are shown in the main text in Figure 2.}
  \label{figNativeXPSFit}
\end{figure}

\begin{figure}
\centering
  \includegraphics[width=\linewidth]{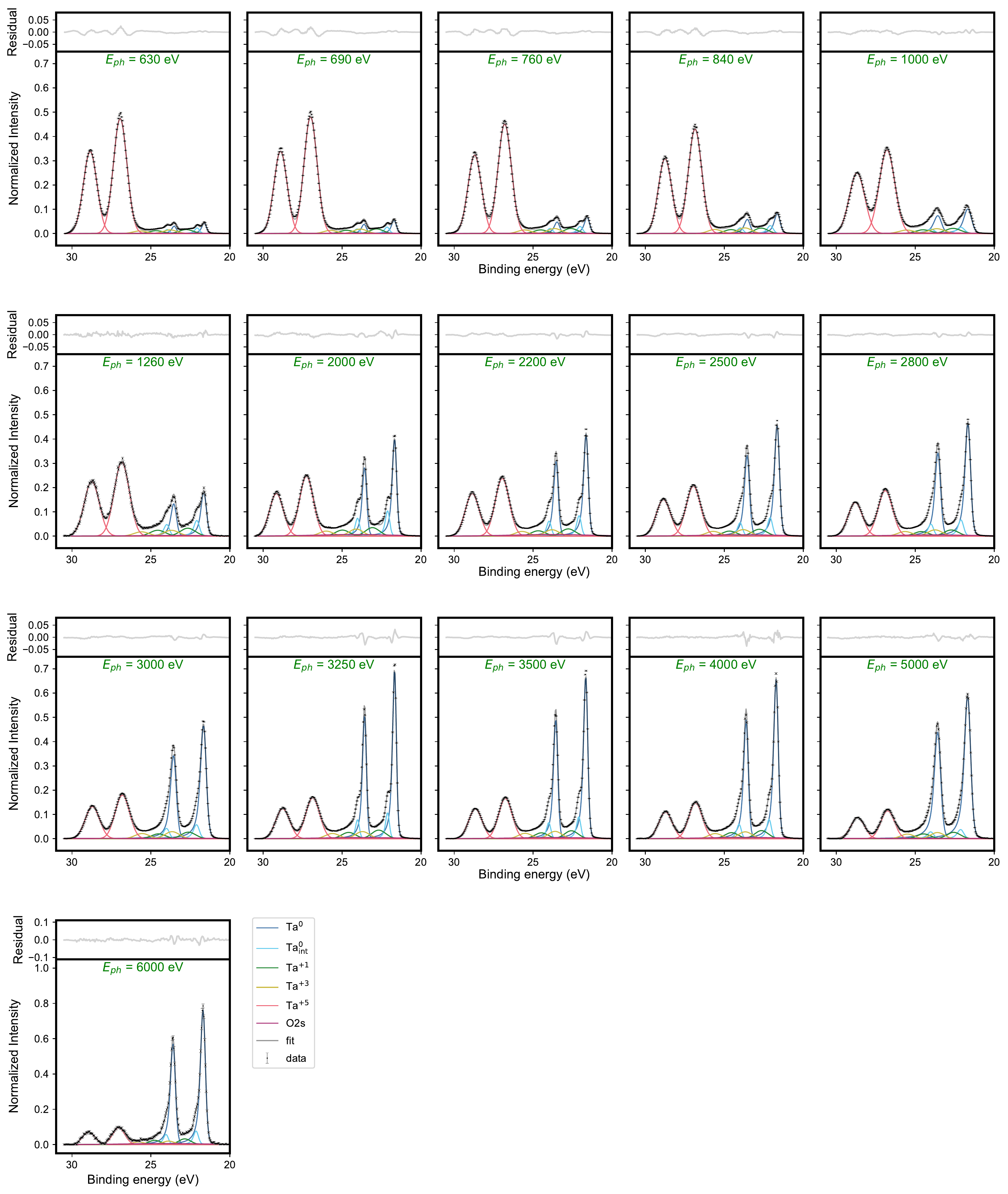}
  \caption{Fitted Ta4f intensity spectra for all X-ray energies on the BOE treated sample. All spectra are fitted simultaneously with certain parameters constrained between spectra, as described in the text. Note that the $E_{ph}=6000$ eV plot has different y-axis limits from the other spectra.}
  \label{figBOEXPSfit}
\end{figure}

\begin{figure}
\centering
  \includegraphics[width=\linewidth]{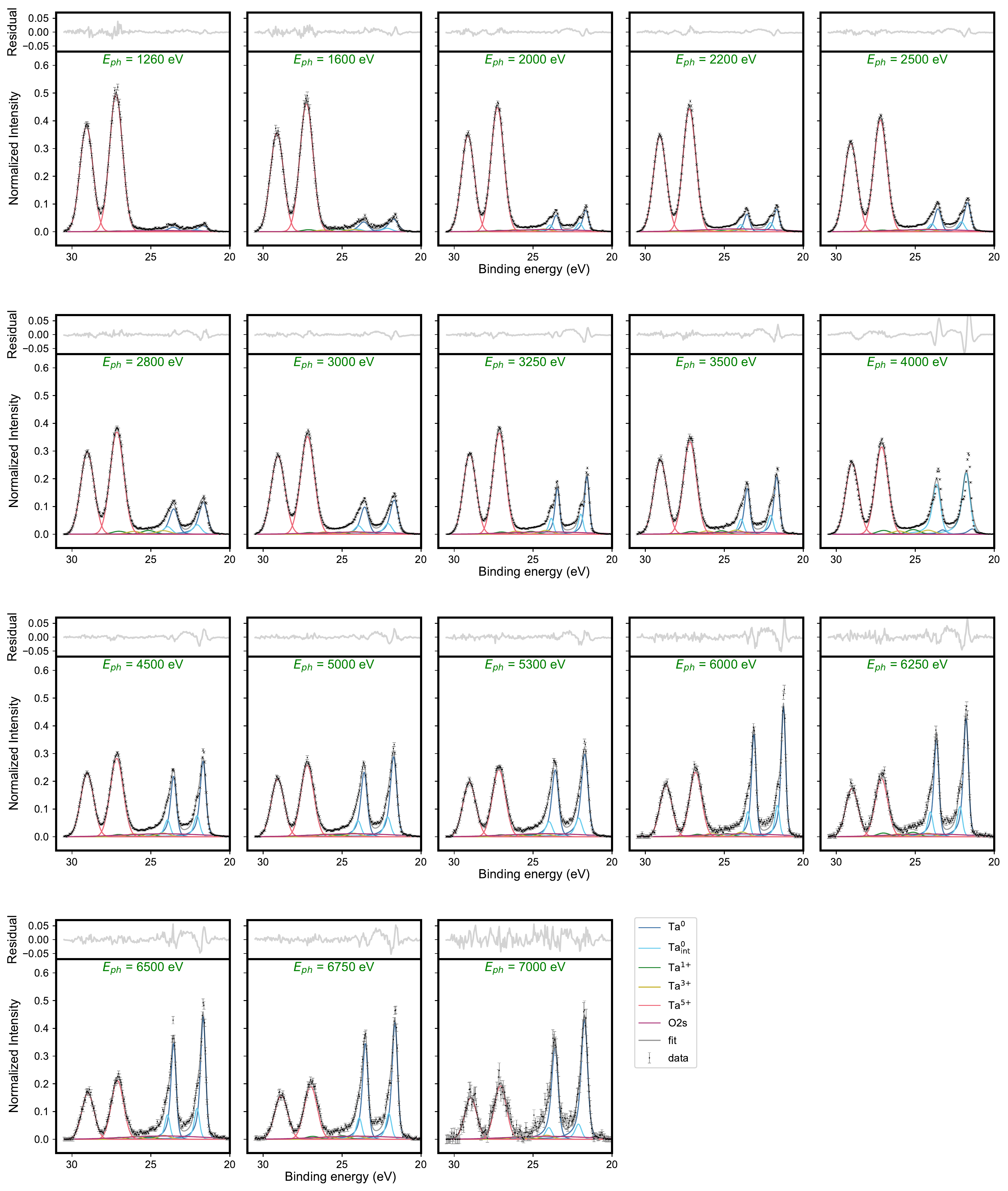}
  \caption{Fitted Ta4f intensity spectra for all X-ray energies on the triacid treated sample. All spectra are fitted simultaneously with certain parameters constrained between spectra, as described in the text.}
  \label{figTriacidXPSfit}
\end{figure}

With these constraints in place, for our native tantalum dataset with 17 different X-ray energies, we have 193 free parameters to fit 187 different peaks. This reduction is a significant improvement over the naive method where each parameter is independent, which requires 340 free parameters for the same 187 peaks. The full results of this fitting method are shown for the native tantalum (Figure \ref{figNativeXPSFit}), BOE treated tantalum (Figure \ref{figBOEXPSfit}), and triacid treated tantalum (Figure \ref{figTriacidXPSfit}).

\subsection{Propagation of error to photoelectron intensity fractions}

The fitted parameters from the XPS spectra which are used in the depth profile analysis are the intensity fractions of each of the Ta$_{7/2}$ peaks, 
\begin{equation} \label{eq:int_fraction}
    W_n = \frac{A_n}{\sum\limits_m A_m},
\end{equation}

where $W_n$ is the intensity fraction of the oxidation state $n$, $A_n$ is the intensity from the oxidation state $n$, and the index $n$ belongs to the set $\{$Ta$^0$, Ta$^0_\text{int}$, Ta$^{1+}$, Ta$^{3+}$, Ta$^{5+}\}$. In addition to $f_n$, we will also need $\sigma_{f_n}$, the uncertainty in the intensity fraction.

We propagate the uncertainty to the intensity fraction using the following formula \cite{bevington_error_2003}:
\begin{equation} \label{eq:error_prop_general}
    \sigma_B^2 = \sum_{i,j} \sigma_{b_ib_j}^2\left( \frac{\partial B}{\partial b_i}\right)\left( \frac{\partial B}{\partial b_j}\right),
\end{equation}

where $B$ is a function of the $b_i$s, and $\sigma_{b_ib_j}^2$ is the covariance between $b_i$ and $b_j$, and $\sigma_{b_ib_i}^2$ =  $\sigma_{b_i}^2$ is the variance of $b_i$. In Equation \ref{eq:error_prop_general}, we have not assumed that the errors are uncorrelated, as the different intensities do correlate with each other.

The empirical covariance matrix is calculated and reported by the lmfit Python module in addition to the fitted parameters. We apply Equation \ref{eq:error_prop_general} to Equation \ref{eq:int_fraction} to arrive at:
\begin{equation}
    \sigma_{W_n}^2 = \sigma_{A_n} \left( \frac{1-W_n}{\sum\limits_m A_m} \right)^2 + \left( \frac{W_n}{\sum\limits_m A_m} \right)^2 \sum\limits_{\ell \neq n}\sigma_{A_\ell}^2 - 2\frac{W_n(1-W_n)}{\bigg(\sum\limits_m A_m\bigg)^2} \sum\limits_{\ell{}\neq n}\sigma_{A_\ell{}A_n}^2,
\end{equation}

which we use to set the uncertainties in the intensity fractions used when fitting a depth profile.

\section{Chemical depth profile analysis}
\subsection{General modeling of the film}

Here we model the object of study as a multicomponent thin film placed on a uniform substrate of infinite thickness. The object occupies the half-space of $x\geq0$, and contains $N$ unique species (${S_n}, n = 1,...,N$) that are spatially mixed, including the substrate species ($S_N$). The mixing is inhomogenous along $x$ but is homogenous along the other two dimensions. During mixing, we assume the volume of each species is conserved, so that a volume fraction profile $\{F_n(x)\}$ $(n =1,...,N)$ is defined for each depth $x$, with the total volume fraction constraint $\sum\limits_{n=1}^N F_n(x) = 1$ for all $x$. The volume fraction of the substrate species, $F_N$, follows the limiting behavior of $\lim_{x\to\infty} F_N(x)=1$. 

Now consider an atom of interest Q, that has a mass density of $\rho_n$ within each species $S_n$. As such, the total mass density of all atoms Q at a depth of $x$ is
\begin{equation}
\rho(x) = \displaystyle\sum_{n=1}^N\rho_nF_n(x).
\label{mass_density}
\end{equation}

\subsection{Photon and photoelectron attenuation}

Consider a beam of incident photons that have photon energy $E_\textrm{ph}$, incident angle $\Theta_\textrm{ph}$, and photon flux $I_0$ normalized to unity. The beam is attenuated across the thin film due to photon absorption by atom Q. Across a thin slab of thickness $ds$ at a depth of $s$, the attenuation of beam flux can be written as    
\begin{equation}
dI = -I(s, E_\textrm{ph}, \Theta_\textrm{ph})\frac{\mu}{\rho_0}\frac{\rho(s)}{\sin\Theta_\textrm{ph}}ds,
\label{xattn_diff}
\end{equation}
in which the term $\mu/\rho_0$ is the tabulated mass attenuation coefficient as a function of $E_\textrm{ph}$, i.e., $\mu/\rho_0=\mu(E_\textrm{ph})/\rho_0$. Integration over Equation (\ref{xattn_diff}) produces photon flux at the depth $x$, as
\begin{equation}
I(x, E_\textrm{ph}, \Theta_\textrm{ph}) = \exp\left(-\int_0^x\frac{\mu(E_\textrm{ph})}{\rho_0}\frac{\rho(s)}{\sin\Theta_\textrm{ph}}ds\right).
\label{xattn}
\end{equation}

For photoelectron generation, we assume an atom Q has a photoelectron yield $\eta$ that is independent of its species assignment. As such, the species $S_n$ contained within a thin slab of $dx$ at a depth of $x$ should give rise to a photoelectron generation rate $dg_n$ as:
\begin{align}
dg_n & = \eta\rho_nF_n(x)I(x, E_\textrm{ph}, \Theta_\textrm{ph})dx  \\
& = \eta\rho_nF_n(x)\exp\left(-\int_0^x\frac{\mu(E_\textrm{ph})}{\rho_0}\frac{\rho(s)}{\sin\Theta_\textrm{ph}}ds\right)dx. \nonumber
\label{pe_gen}
\end{align}
Photoelectrons generated from atoms Q in species $S_n$ have kinetic energy $E_k = E_\textrm{ph}-E_{b,n}$, in which $E_{b,n}$ is the binding energy of the core level of Q in $S_n$. With a photoelectron collection angle $\Theta_\textrm{el}$, photoelectrons generated at depth $x$ travel through a distance of $x/\sin\Theta_\textrm{el}$ across the film, leading to an attenuation factor of $\exp\left(-x\lambda_\textrm{el}^{-1}(E_\textrm{ph}-E_{b,n})/\sin\Theta_\textrm{el}\right)$, in which $\lambda_\textrm{el}(E_\textrm{ph}-E_{b,n})$ is the inelastic mean free path of a photoelectron at the energy $E_\textrm{ph}-E_{b,n}$. As such, the species $S_n$ contained within a thin slab of $dx$ at a depth of $x$ will contribute 
\begin{align}
dA_n & =\gamma dg_n\exp\left(-\frac{x}{\lambda_\textrm{el}(E_\textrm{ph}-E_{b,n})\sin\Theta_\textrm{el}}\right)  \\
& = \gamma\eta\rho_nF_n(x)\exp\left(-\int_0^x\frac{\mu(E_\textrm{ph})}{\rho_0}\frac{\rho(s)}{\sin\Theta_\textrm{ph}}ds-\frac{x}{\lambda_\textrm{el}(E_\textrm{ph}-E_{b,n})\sin\Theta_\textrm{el}}\right)dx \nonumber \\
& = \gamma\eta\rho_nF_n(x)T_n(x, E_\textrm{ph},\Theta_\textrm{ph},\Theta_\textrm{el})dx \nonumber
\label{pe_collect}
\end{align}
to the final photoelectron spectrum, where $\gamma$ is the photoelectron collection efficiency and $T_n(x, E_\textrm{ph},\Theta_\textrm{ph},\allowbreak \Theta_\textrm{el})$ covers the exponential attenuation term in the second line. It should be noted that the latter term only has very weak dependence on $n$, as the variance of $E_{b,n}$ has little relative influence on photoelectron kinetic energy $E_\textrm{el}-E_{b,n}$. As such, we may safely replace $T_n(x, E_\textrm{ph},\Theta_\textrm{ph},\Theta_\textrm{el})$ with a species-independent attenuation $T(x, E_\textrm{ph}, E_b, \Theta_\textrm{ph},\Theta_\textrm{el})$, in which $E_b\simeq\langle E_{b,n}\rangle$ is the typical binding energy.

Integration over $x$ produces the total contribution of species $S_n$ to the final spectrum, as
\begin{equation}
A_n(E_\textrm{ph}, E_b, \Theta_\textrm{ph},\Theta_\textrm{el}) =  \gamma\eta\rho_n\int_0^\infty F_n(x)T(x, E_\textrm{ph}, E_b, \Theta_\textrm{ph},\Theta_\textrm{el})dx.
\label{pe_collect_2}
\end{equation}
Finally, the species $S_n$ contributes a normalized weight of
\begin{equation}
W_n(E_\textrm{ph}, E_b, \Theta_\textrm{ph},\Theta_\textrm{el}) =  \frac{\rho_n\displaystyle\int_0^\infty F_n(x)T(x, E_\textrm{ph}, E_b, \Theta_\textrm{ph},\Theta_\textrm{el})dx}{\displaystyle\sum_{n=1}^{N}\rho_n\displaystyle\int_0^\infty F_n(x)T(x, E_\textrm{ph}, E_b, \Theta_\textrm{ph},\Theta_\textrm{el})dx},
\label{XPS_spectral_weight}
\end{equation}
which directly corresponds to the results obtained by analyzing experimental XPS spectra. The fitting process will use experimentally obtained $\{W_n(E_\textrm{ph}, E_b, \Theta_\textrm{ph},\Theta_\textrm{el})\}$ to infer volume fractions $\{F_n(x)\}$.

\subsection{Basis set of volume fractions}
\label{Basis set of volume fractions}
\begin{figure}
\centering
  \includegraphics[width=0.7\linewidth]{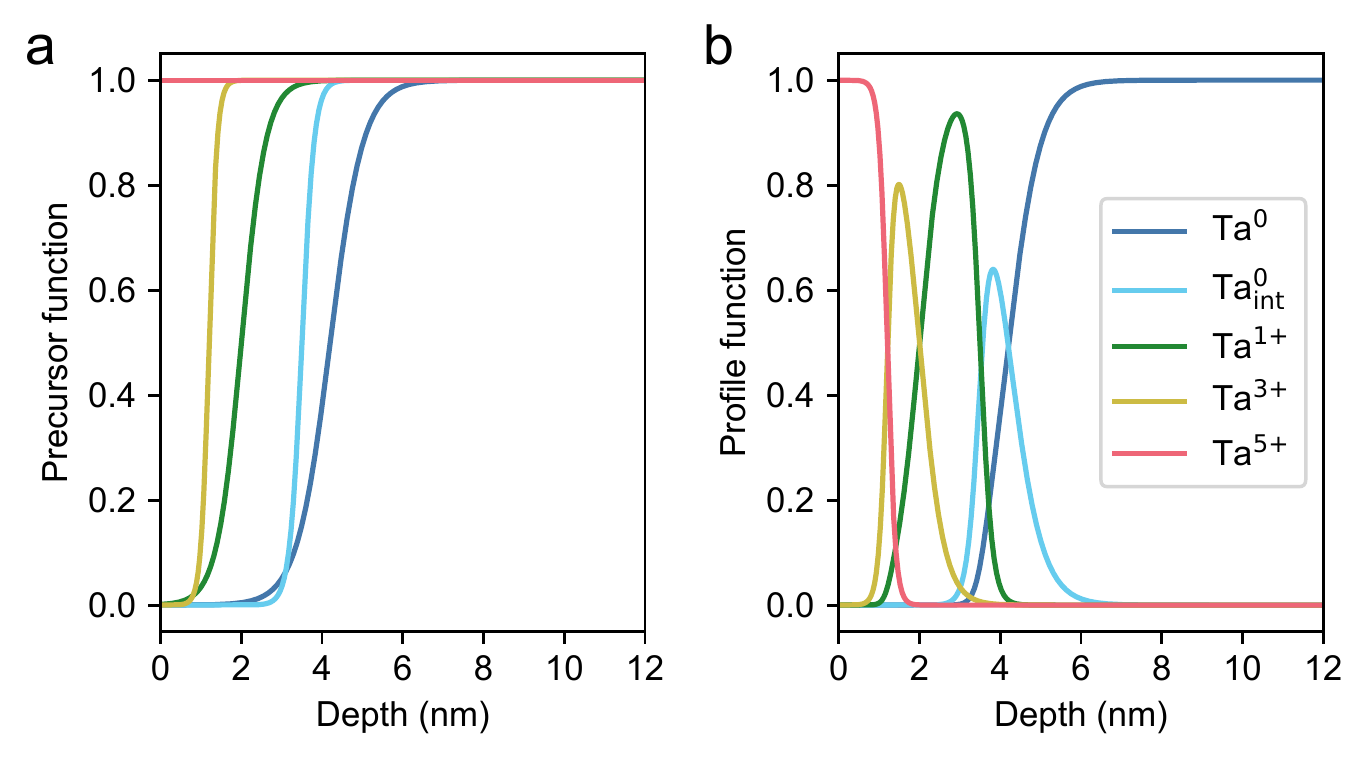}
  \caption{Example precursor function set (a) and corresponding profile function set (b). The precursor functions are sigmoids or a constant as described in Equation \ref{precursors}, and the profile functions are calculated from the precursors according to Equation \ref{volFrac}. The profile function set has the property that at any depth, the sum of all functions is unity.}
  \label{fig:sup_precursor}
\end{figure}

A basis set is needed for volume fractions $\{F_n(x)\}$, so that a small number of fitting parameters can be used to model the spatial distribution of each species within the thin film. The basis set we use is generated with the following procedure. First, a set of $N$ \emph{precursor functions} are defined as 
\begin{equation}
f_n(x) =
\begin{cases}
H(x) & (n=1)\\
\left[1 + \exp\left(\frac{d_n - x}{w_n}\right)\right]^{-1} &  (1<n\leq N),
\end{cases}
\label{precursors}
\end{equation}
in which $H(x)$ is the unit step function, and $d_n$ and $w_n$ are center and width parameters for $f_n(x)$. Using the precursor functions, the volume fraction functions are generated as
\begin{equation}
F_n(x) =
\begin{cases}
\left(1-f_{n+1}(x)\right)\displaystyle\prod_{i=1}^n f_i(x)& (1\leq n<N)\\
\displaystyle\prod_{i=1}^N f_i(x) &  (n = N).
\end{cases}
\label{volFrac}
\end{equation}
Equation \ref{volFrac} will generate functions which automatically follow the sum rule $\sum\limits_{n=1}^N F_n(x) = 1$. A typical initial guess of precursor functions $\{f_n(x)\}$ and generated volume fractions $\{F_n(x)\}$ are shown in Fig. \ref{fig:sup_precursor}. The basis set for the $N$ species has a total of $2(N-1)$ fitting parameters ($N-1$ $d_n$'s and $N-1$ $w_n$). Note that the precursor function $f_1(x)$ requires no parameters. With $N=5$ unique species, the fitting procedure involves 8 fitting parameters.

\subsection{Method to ensure a best fit}

Fitting the model function, Equation \ref{XPS_spectral_weight}, with the depth profile parameterizations, described in Section \ref{Basis set of volume fractions}, to experimental data is a non-convex optimization problem. A fitting algorithm can terminate in a local minimum of the least squared objective function instead of a global minimum. Unlike the fits to XPS spectra described in Section \ref{XPS_trace_analysis}, we may not have sufficient physical insight into our system to recognize and discard solutions corresponding to these local minima when we fit chemical depth profiles.

\begin{figure}
\centering
  \includegraphics[width=1\linewidth]{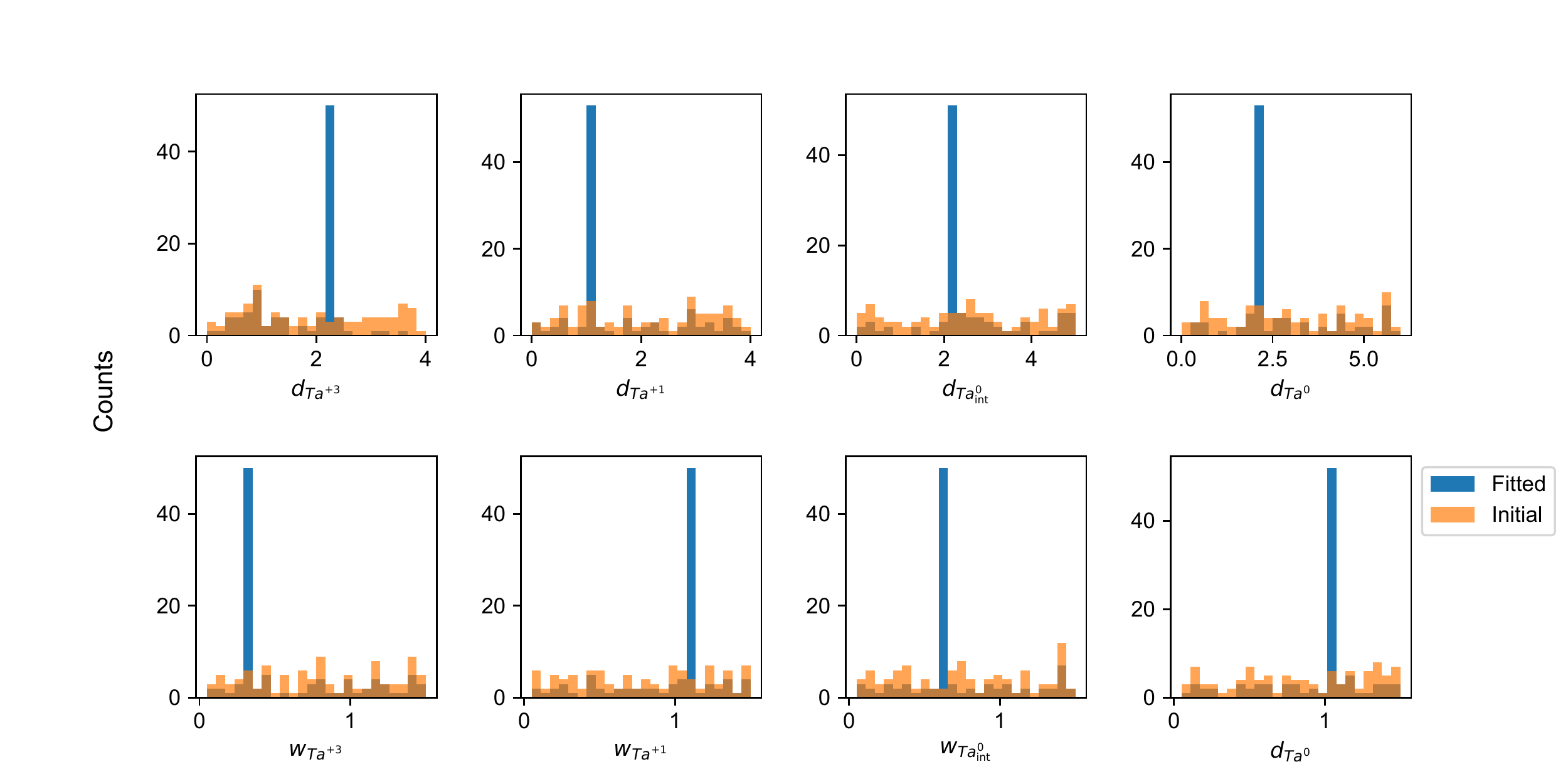}
  \caption{Initial and final parameters from 100 fits to the native oxide sample. Initial conditions for each fit were drawn from uncorrelated uniform distributions with bounds given in Table \ref{table:sup_bounds}. Parameters are those specified in Equation \ref{precursors}, in units of nm, with numeric subscripts $\{1, 2, 3, 4\}$ replaced by the descriptive subscripts $\text{Ta}^{5+}, \text{Ta}^{3+}, \text{Ta}^{1+} \{\text{Ta}^0_{\text{int}}\}$}
  \label{fig:sup_MonteCarloFit}
\end{figure}

To overcome this issue, we perform 100 fits to the data from each sample, varying the initial conditions for each fit. For each fit, we draw initial conditions for the 8 parameters (4 $d_n$s and 4 $w_n$s specified in Equation \ref{precursors}) from uncorrelated uniform distributions. The bounds for these uniform distributions are given in Table \ref{table:sup_bounds}; the bounds for the triacid treated sample were extended to account for the thicker oxide layer. For each sample we fit, we found that a large fraction of the 100 fits converged to the same set of final parameters. These sets of fitted parameters correspond to the lowest $\chi^2$ value, and therefore we are confident that we have found the best chemical depth profile fit given our parameterization. Histograms of the initial and final parameter values for all 100 fits to the data taken from the native sample are shown in Figure \ref{fig:sup_MonteCarloFit}.

\begin{table}[h]
 \caption{Uniform distribution bounds used for the initial conditions of depth profile fits. Parameters are the $d_n$s and $w_n$s from Equation \ref{precursors} with numeric subscripts $\{2, 3, 4, 5\}$ replaced by the descriptive subscripts $\text{Ta}^{3+}, \text{Ta}^{1+}, \{\text{Ta}^0_{\text{int}}, \text{Ta}^{0}\}$. All values are given in nm.} 
 \label{table:sup_bounds}
 \centering
 \setlength{\tabcolsep}{10pt}
  \begin{tabular}[htbp]{@{}lcccccccc@{}}
    \hline
    Film & $d_{\text{Ta}^{3+}}$ & $d_{\text{Ta}^{1+}}$ & $d_{\text{Ta}^0_{\text{int}}}$ & $d_{\text{Ta}^{0}}$
     & $w_{\text{Ta}^{3+}}$ & $w_{\text{Ta}^{1+}}$ & $w_{\text{Ta}^0_{\text{int}}}$ & $w_{\text{Ta}^{0}}$ \\
    \hline
    Samples not triacid treated & [0, 3] & [0, 4] & [0, 5] & [0, 6] & [0.05, 1.5] & [0.05, 1.5] & [0.05, 1.5] & [0.05, 1.5] \\
    Triacid treated sample & [2, 6] & [0, 6] & [0, 7] & [0, 8] & [0.05, 1.5] & [1.05, 2.5] & [1.05, 2.5] & [2.05, 3.5] \\
    \hline
    
  \end{tabular}
\end{table}

\subsection{Uncertainties in effective thickness}

Uncertainties for the effective thicknesses of surface species are given in Table 1 in the main text and Table \ref{table:sup_multipleBOE}. As the effective thickness is not a parameter that is fit in the model, we calculate these uncertainties using a post-fit Monte-Carlo method.

For each sample, we perturb the set of best fit parameters by adding a set of numbers drawn from uncorrelated uniform distributions. We either accept or reject the perturbed parameters based on the corresponding $\chi^2$ value. After we have accepted 100 perturbed parameters, we take the standard deviation of the set of 100 calculated effective thicknesses to report the 1$\sigma$ uncertainty.

The criteria for accepting a perturbed set of parameters is whether, given our data, the likelihood of the parameters exceeds 0.3. We calculate likelihoods from each parameter set from the $\chi^2$ distribution for our fit; $\chi^2$ values and probabilities are scaled such that the best fit has the maximum probability at 1.

\subsection{Results of fitting algorithm}

\begin{figure}
\centering
  \includegraphics[width=0.7\linewidth]{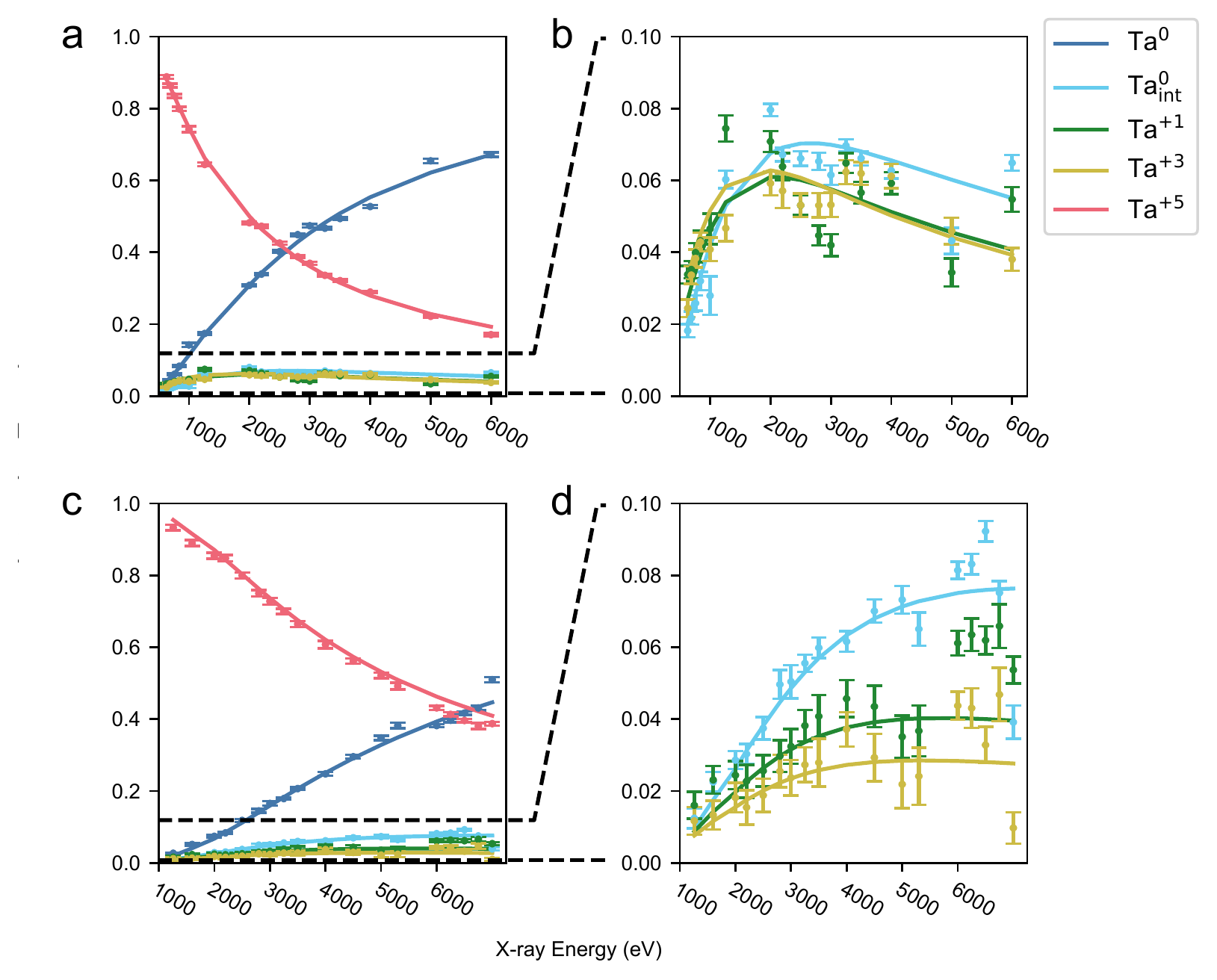}
  \caption{Experimental and simulated photoelectron intensities for the BOE treated (a-b) and triacid treated (c-d) samples. These fits correspond to the depth profiles shown in the main text in Figure 4b and Figure 4c respectively.}
  \label{fig_BOE_triacid_fit}
\end{figure}

The fitted photoelectron intensity fractions for the untreated (``Native") tantalum are shown in the main text in Figure 3, and the depth profile is shown in Figure 4a. Depth profiles for the BOE treated and triacid treated tantalum samples are shown in the main text in Figure 4b and Figure 4c, respectively. The fitted photoelectron intensity fractions for these latter two samples are shown in Figure \ref{fig_BOE_triacid_fit}.

\section{Peak assignment} \label{sec:peak_assignment}

As described in the main text, we observe 5 Ta4f doublets. The pair of peaks at 22 eV and 24 eV binding energy has previously been assigned as Ta$^{0}$, the pair at 27 eV and 29 eV assigned to Ta$^{5+}$, and the pair at the shoulder of the Ta$^0$ peaks assigned as a tantalum species at a material interface with a differing coordination number \cite{Himpsel_core_level_shift:1984}. The Ta$^{5+}$ in our sample is in the form Ta$_2$O$_5$. The other two doublets have binding energies at 23 eV and 25 eV, and 24 eV and 26 eV. We assign these two intermediate doublets as Ta$^{1+}$ and Ta$^{3+}$ based on their similarity to the peak locations reported in \cite{Himpsel_core_level_shift:1984}.

The binding energy position of the Ta$^{1+}$ and Ta$^{3+}$ peaks indicates their charge state, but does not indicate to which chemical compound they belong. A wide survey scan of the untreated (''Native") sample did not indicate any major elements other than tantalum, oxygen, and carbon. We wanted to rule out the presence of nitrides, so we performed a fine scan on the nitrogen KLL Auger line and did not see a peak. The usual line for nitrogen, N1s, overlaps with the Ta4p$_{3/2}$ line. 

\begin{figure}[htbp]
  \centering
  \includegraphics[width=0.8\linewidth]{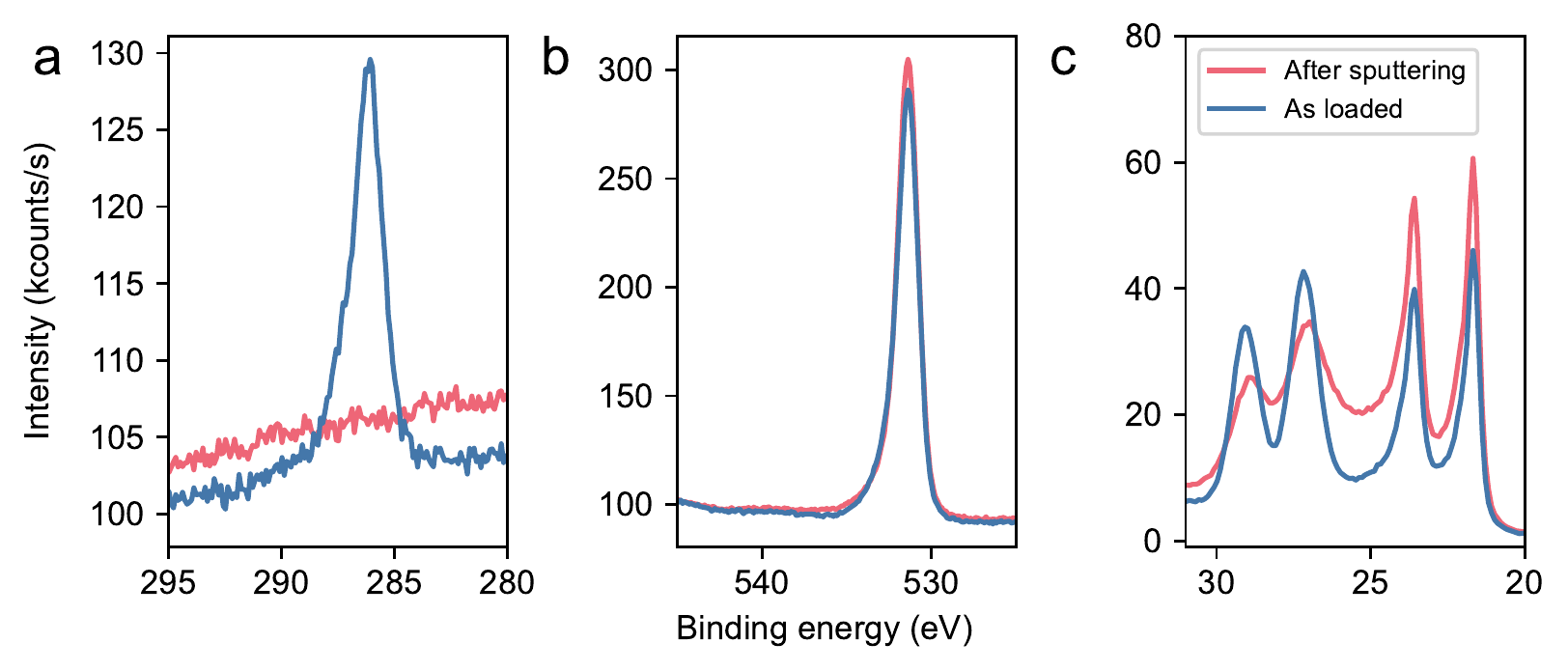}
  \caption{a) C1s spectra for a tantalum film before and after Ar ion milling. The carbon signal is gone after milling. b) O1s spectra for a tantalum film before and after Ar ion milling. Little change is observed. c) Ta4f spectra of tantalum film before and after Ar ion milling. Based on tool calibration, we approximately etched away 1 nm of Ta$_2$O$_5$. Some Ta$^{5+}$ signal remains after milling. The photoelectron contribution from tantalum oxidation state(s) with binding energies between the Ta$^{5+}$ and Ta$^{0}$ peaks remain. These results indicate that these intermediate binding energy compounds are not carbides.}
  \label{fig:sup_milling}
\end{figure}

To rule out that the Ta$^{1+}$ and Ta$^{3+}$ peaks are carbides, we performed an experiment on a separate film using a ThermoFisher K-Alpha X-Ray Photoelectron Spectrometer with an Ar$^+$ ion gun. We scanned the Ta4f and C1s peaks, then sputtered the film, and then scanned again. The manufacturer provided a calibrated rate curve for sputtering Ta$_2$O$_5$, from which we estimate our etch removed approximately 1 nm from the surface of the film. Before sputtering, we observed similar C1s, O1s, and Ta4f spectra to what we observed in the VEXPS dataset at similar X-ray energies. After sputtering, we observe no C1s peak, little change to the O1s peak, a smaller Ta$^{5+}$ peak, non-zero intensity between the Ta$^0$ and Ta$^{5+}$ peaks (the regions in which the Ta$^{1+}$ and Ta$^{3+}$ peaks are located). These results are shown in Figure \ref{fig:sup_milling}. We conclude that the Ta$^{1+}$ and Ta$^{3+}$ oxidation states are in the forms of amorphous Ta$_2$O and amorphous Ta$_2$O$_3$. 

\section{Effect of piranha clean}  \label{sec:piranha}
All samples underwent piranha cleaning (outlined in Section \ref{sec:methods}) before VEXPS measurement or further chemical processing, matching the process used in \cite{kevin_resonator}, where all devices were piranha cleaned prior to measurement or surface processing. To characterize the effect this piranha clean has on the tantalum surface, we performed XPS with a ThermoFisher K-Alpha X-Ray Photoelectron Spectrometer on a tantalum film before and immediately after a piranha clean, as well as after a subsequent 20 minute BOE treatment, matching the processing of the ``BOE" surface studied in the main text.

\begin{figure}[htbp]
\centering
  \includegraphics[width=0.9\linewidth]{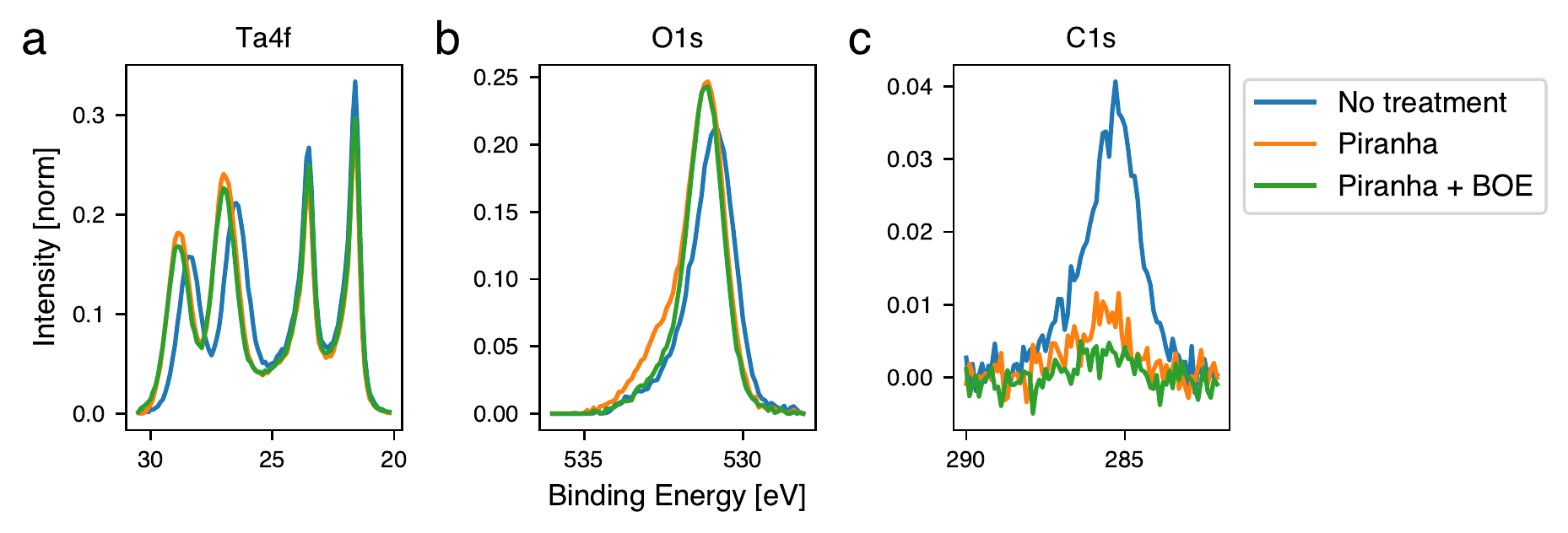}
  \caption{XPS measurements of the Ta4f (a), O1s (b), and C1s (c) spectra, taken on a tantalum sample before piranha cleaning, after piranha cleaning (``native" surface), and after both piranha cleaning and a 20 minute BOE treatment (``BOE" surface). Ta4f and O1s data are Shirley background corrected \cite{engelhard_introductory_2020} and the C1s spectra have linear backgrounds subtracted. All spectra on a sample are normalized so the total intensity under the Ta4f spectrum for that sample is unity.}
  \label{fig:sup_piranha}
\end{figure}

In a wide survey scan, we observed a small Na1s peak, but this peak disappears after the piranha clean. The Ta4f spectrum shows a small increase in the intensity of the Ta$^{5{+}}$ doublet after piranha cleaning, followed by a decrease after the BOE treatment (Figure \ref{fig:sup_piranha}(a)). We also observe an increase in the O1s spectrum intensity after piranha and a decrease after the BOE treatment (Figure \ref{fig:sup_piranha}(b)). The C1s spectrum, by contrast, shows a marked decrease in intensity after piranha, and a further decrease after BOE treatment (Figure \ref{fig:sup_piranha}(c)).

We attribute the Na1s peak seen in the large survey scan to contamination in the lab. Both the Ta4f and O1s spectra indicate a small increase in the oxide thickness after piranha cleaning, and the measurements taken after the 20 minute BOE treatment indicate that the oxide thickness is still larger than when the film was freshly sputtered. The intensity of the C1s peak shows the effectiveness of the piranha clean at removing hydrocarbons from the surface of the sample. The further decrease in C1s intensity after the BOE treatment may be due to the hydrocarbons lifting off as the oxide is etched. We note that the C1s peak intensity in this measurement is generally larger than those measured with VEXPS; as the carbon is entirely adventitious (Section \ref{sec:peak_assignment}), variations in the environment and time between cleaning and measurement could have a significant impact on the signal.

\section{Effect of multiple BOE treatments}
\label{Effect of multiple BOE treatments}
In Figure 4 in the main text, we show that the BOE treatment affectw not just the Ta$^{+5}$, but also the Ta$^{+1}$, Ta$^{+3}$, and Ta$^{0}_\text{int}$ species, which do not appear to be exposed on the surface of the material. One hypothesis that can explain how a BOE surface treatment affects the interface species between the Ta$^0$ and Ta$^{5+}$ species is that the BOE treatment strips away all of the Ta$_2$O$_5$, interacts with the underlying layers, and then the Ta$_2$O$_5$ layer grows back when the sample is exposed to air after the treatment.

\begin{figure}
\centering
  \includegraphics[width=0.7\linewidth]{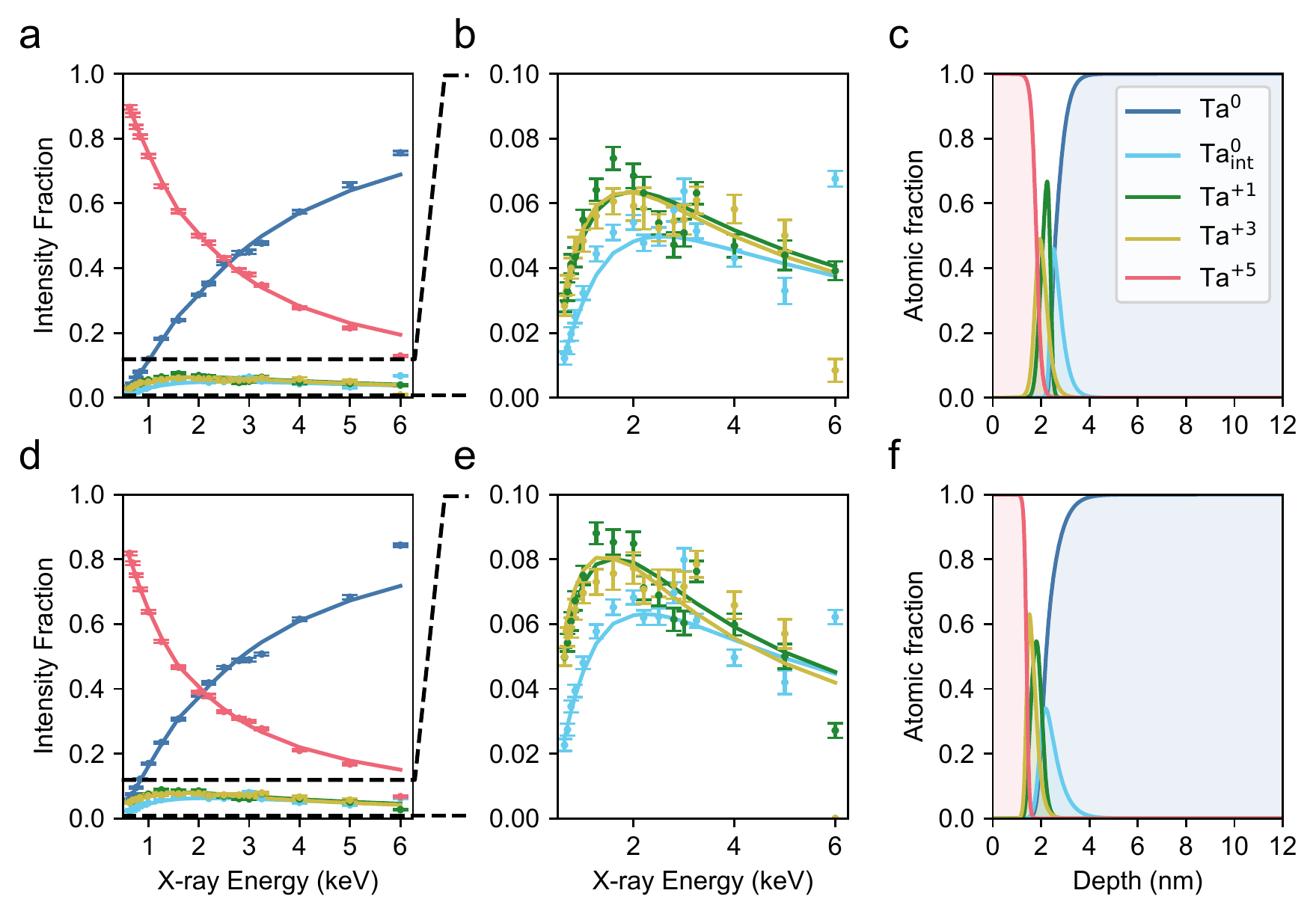}
  \caption{Depth profile fit results for the BOE-1x and BOE-2x samples. a-b) Experimental and simulated relative photoelectron intensities as a function of incident X-ray energy for BOE-1x. c) Fitted interface depth profile for BOE-1x. d-f) as a-c, but for BOE-2x.}
  \label{fig:sup_BOE2x}
\end{figure}

To test this hypothesis, we performed VEXPS on two samples from a tantalum film. Both samples were treated two weeks before the VEXPS measurement, and one was treated again in BOE immediately before the VEXPS measurements. We denote these two samples as BOE-1x and BOE-2x, respectively. We analyzed the VEXPS data from these samples in the same manner as the native, BOE treated, and triacid treated samples in the main text. The results of this analysis are shown in Figure \ref{fig:sup_BOE2x}, and the effective thickness of the Ta$^{5+}$, Ta$^{3+}$, Ta$^{1+}$, and Ta$^{0}_\text{int}$ species are shown along with those of the native, BOE treated, and triacid treated samples from the main text in Table \ref{table:sup_multipleBOE}.

The Ta$^{5+}$ thickness of the BOE-1x samples and the BOE treated sample from the main text are similar, both being approximately 0.4 nm smaller than that of the native sample. The Ta$^{5+}$ thickness of the BOE-2x sample is decreased by approximately another 0.4 nm from that of the singly BOE treated samples. As the effect of etching in BOE is roughly additive, we conclude that the BOE treatment is not completely etching away the Ta$_2$O$_5$.

\begin{table}[h]
 \caption{Effective thickness of different tantalum oxidation states as obtained from depth profile fitting for different tantalum films. All data in nm. Uncertainties are $\pm{}1\sigma{}$ confidence intervals reflecting uncertainty in the fit.}
 \label{table:sup_multipleBOE}
 \centering
 \setlength{\tabcolsep}{10pt}
  \begin{tabular}[htbp]{@{}lllll@{}}
    \hline
    Film & Ta$^{5+}$ & Ta$^{3+}$ & Ta$^{1+}$ & Ta$^{0}_{\text{int}}$ \\
    \hline
    Native (main text) & 2.257 $\pm$ 0.023 & 0.370 $\pm$ 0.016 & 0.370 $\pm$ 0.017 & 0.368 $\pm$ 0.019 \\
    BOE (main text) & 1.853 $\pm$ 0.028 & 0.296 $\pm$ 0.022 & 0.302 $\pm$ 0.023 & 0.400 $\pm$ 0.021\\
    Triacid (main text) & 4.826 $\pm$ 0.036 & 0.379 $\pm$ 0.016 & 0.545 $\pm$ 0.020 & 1.198 $\pm$ 0.027\\
    BOE-1x & 1.824 $\pm$  0.017 & 0.283 $\pm$ 0.017 & 0.285 $\pm$ 0.014 & 0.382 $\pm$ 0.020\\
    BOE-2x & 1.430 $\pm$ 0.014 & 0.294 $\pm$ 0.017 & 0.314 $\pm$ 0.017 & 0.302 $\pm$ 0.019\\
    \hline

  \end{tabular}
\end{table}

\section{Pinholes measured by atomic force microscopy}
\label{AFM_section}
Figure 4 in the main text shows the BOE treatment affecting the Ta$^{3+}$, Ta$^{1+}$, and Ta$^{0}_\text{int}$ species, even though these species are not at the surface of the samples. Based on the results in Section \ref{Effect of multiple BOE treatments}, we do not believe that the BOE treatment is removing the Ta$^{5+}$ layer and affecting the underlying layers.

\begin{figure}
    \centering
    \includegraphics[width=0.8\linewidth]{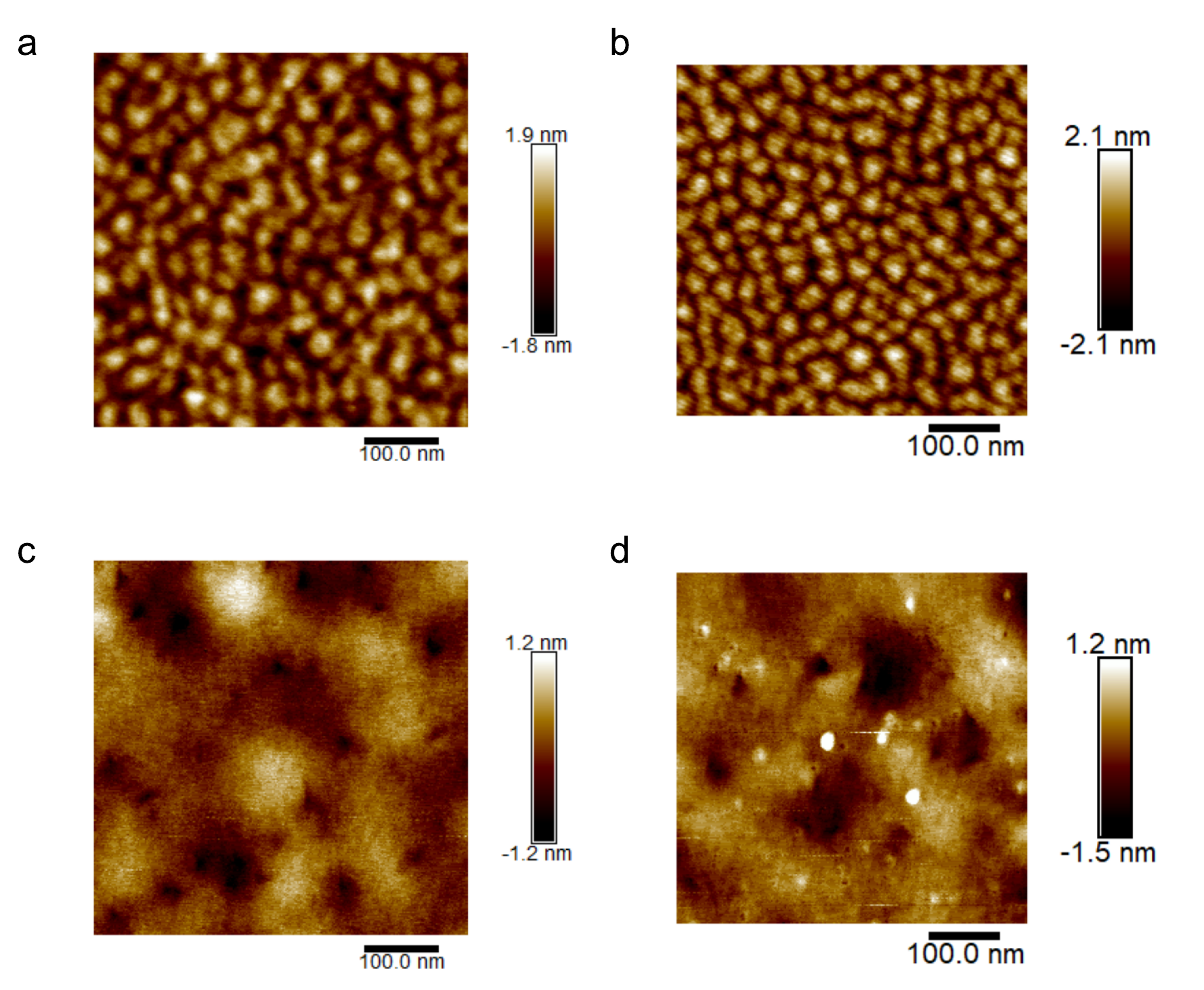}
    \caption{Atomic force microscopy (AFM) image of the height of tantalum samples. a) Untreated sample from the same film as the native, BOE treated, and triacid treated samples described in the main text with extracted root mean square roughness of 0.568 nm. b) as a), but BOE treated with extracted root mean square roughness of 0.648 nm. c) Untreated sample from an identically deposited film as the BOE-1x and BOE-2x samples described in Section \ref{Effect of multiple BOE treatments}, with extracted root mean square roughness of 0.324 nm. d) As c), but BOE treated with extracted root mean square roughness of 0.383 nm.}
  \label{fig:sup_AFM}
\end{figure}

We performed atomic force microscopy (AFM) on untreated films using a Bruker ICON3 Atomic Force Microscope. We performed AFM on samples from both the film as the three samples (native, BOE treated, and triacid treated) that were discussed in the main text and a film deposited with the same conditions as the film which the BOE-1x and BOE-2x samples. While untreated samples from both films show noticeably different surface morphologies, the measured surface roughnesses over 500 nm squares are significant compared to the 2.257 nm $\pm$ 0.023 nm thick Ta$^{5+}$ layer we found on our native oxide film with VEXPS (Figure \ref{fig:sup_AFM}. We hypothesize that the observed uneven surface morphologies allow the BOE solution access to the buried interface. We note that the observed surface morphologies do not qualitatively change after treatment in BOE.

We must interpret our fitted depth profiles given the observed surface roughness. The XPS spectra were measured with an X-ray beam area of approximately 47 000 $\mu$m$^2$. This area is far larger than the size of features we resolve in either panel of Figure \ref{fig:sup_AFM}. We interpret our fitted depth profiles as an average species fraction through the depth, with the $x$ = 0 depth corresponding to the mean height of our samples.

\section{Disclaimer}
Certain commercial equipment, instruments, or materials are identified in this paper in order to specify the experimental procedure adequately, and do not represent an endorsement by the National Institute of Standards and Technology.

% References
\medskip

% Use the following code if you wish to generate your bibliography with BibTeX;
% replace the string "MSP-template" below with the name(s) of
% the BibTeX data base(s) you want to use.
% The resulting bibliography-output (the content of the .bbl file)
% must be pasted back into this file before submission.
% Please also include your BibTeX data base file(s) in your submission
% so that we can re-run BibTeX if necessary.
%
\bibliographystyle{MSP}
%\bibliography{MSP-template}

\newpage
\bibliography{references}